\documentclass[letterpaper,twocolumn,accepted=2021-08-08]{quantumarticle}
\pdfoutput=1

\usepackage[utf8]{inputenc}
\usepackage[english]{babel}
\usepackage[T1]{fontenc}

\usepackage[numbers,sort&compress]{natbib}

\usepackage{amsmath}
\usepackage{braket}
\usepackage[dvipsnames]{xcolor}
\usepackage[super]{nth}
\usepackage{graphicx}
\usepackage{amssymb}
\usepackage{hyperref}
\usepackage{xspace}
\usepackage{soul} %

\usepackage{tikz}
\usepackage{qcircuit}
\usepackage{mathtools}

\usepackage{array}
\newcolumntype{x}[1]{%
>{\centering\hspace{0pt}}p{#1}}%

\newcommand{\HadamardTest}{Hadamard Test\xspace} %
\newcommand{\SwapTest}{Swap Test\xspace}
\newcommand{\HTlong}{Entanglement Spectroscopy Hadamard Test\xspace}
\newcommand{\HT}{HT\xspace}

\newcommand{\qeHTlong}{qubit-efficient HT\xspace}
\newcommand{\qeHT}{qe-HT\xspace}
\newcommand{\qeHTfourk}{$4k$ qe-HT\xspace}
\newcommand{\qeHTthreek}{$3k$ qe-HT\xspace}
\newcommand{\qeHTfourklong}{$4k$ \qeHTlong{}\xspace}
\newcommand{\qeHTthreeklong}{$3k$ \qeHTlong{}\xspace}

\newcommand{\TwoCopyTest}{Two-Copy Test\xspace} %

\newcommand{\TCTlong}{Entanglement Spectroscopy Two-Copy Test\xspace}
\newcommand{\TCT}{TCT\xspace}

\newcommand{\qeTCTlong}{qubit-efficient TCT\xspace}
\newcommand{\qeTCT}{qe-TCT\xspace} 
\newcommand{\qeTCTsixk}{$6k$ qe-TCT\xspace} 
\newcommand{\qeTCTfourk}{$4k$ qe-TCT\xspace}
\newcommand{\qeTCTsixklong}{$6k$ \qeTCTlong{}\xspace} 
\newcommand{\qeTCTfourklong}{$4k$ \qeTCTlong{}\xspace}

\newcommand{\abs}[1]{\left\lvert#1\right\rvert}

\newcommand{\tr}{\operatorname{Tr}}
\newcommand{\ketbra}[2]{\ket{#1}\!\bra{#2}}

\newcommand{\SWAP}{\textsc{SWAP}}
\newcommand{\CSWAP}{\textsc{CSWAP}}
\newcommand{\CNOT}{\textsc{CNOT}}

\newcommand{\tsp}{\ensuremath{T_{\text{sp} }}}
\newcommand{\tcs}{\ensuremath{T_{\text{cs} }}}
\newcommand{\tcn}{\ensuremath{T_{\text{cn} }}}
\newcommand{\thad}{\ensuremath{T_{H} }}

\newlength{\mylength}
\begin{document}

\title{Qubit-efficient entanglement spectroscopy using qubit resets}

\author{Justin Yirka}
\email{yirka@utexas.edu}
\affiliation{Department of Computer Science, The University of Texas at Austin, Austin, TX 78712, USA}
\orcid{0000-0001-6173-2465}

\author{Yi\u{g}it Suba\c{s}\i}
\email{ysubasi@lanl.gov}
\affiliation{Computer, Computational, and Statistical Sciences Division, Los Alamos National Laboratory, Los Alamos, NM 87545, USA}
\orcid{0000-0003-1167-6527}

\begin{abstract}
	One strategy to fit larger problems on NISQ devices is to exploit a tradeoff between circuit width and circuit depth. Unfortunately, this tradeoff still limits the size of tractable problems since the increased depth is often not realizable before noise dominates.
    Here, we develop \emph{qubit-efficient} quantum algorithms for entanglement spectroscopy which avoid this tradeoff.
    In particular, we develop algorithms for computing the trace of the $n$-th power of the density operator of a quantum system, $\tr(\rho^n)$, (related to the R\'enyi entropy of order $n$) that use fewer qubits than any previous efficient algorithm while achieving similar performance in the presence of noise, thus enabling spectroscopy of larger quantum systems on NISQ devices.
    Our algorithms, which require a number of qubits independent of $n$, are variants of previous algorithms with width proportional to $n$, an asymptotic difference.
    The crucial ingredient in these new algorithms is the ability to measure and reinitialize subsets of qubits in the course of the computation, allowing us to reuse qubits and increase the circuit depth without suffering the usual noisy consequences.
    We also introduce the notion of \emph{effective circuit depth} as a generalization of standard circuit depth suitable for circuits with qubit resets. 
    This tool helps explain the noise-resilience of our qubit-efficient algorithms and should aid in designing future algorithms.
    We perform numerical simulations to compare our algorithms to the original variants and show they perform similarly when subjected to noise. Additionally, we experimentally implement one of our qubit-efficient algorithms on the Honeywell System Model H0, estimating $\tr(\rho^n)$ for larger $n$ than possible with previous algorithms.
\end{abstract}

\maketitle

\section{Introduction}
\label{sec:introduction}

Full-scale fault-tolerant quantum computers offer eventual advantages over classical computation for a variety of tasks.
While work continues toward such devices, more research is needed on how to utilize near-term devices.
As we develop applications for noisy intermediate-scale quantum (NISQ) computers~\cite{Preskill2018quantum,IBM_NISQ2019}, a primary limitation is the inverse relationship between the quality and the quantity of available qubits, i.e. larger devices tend to be noisier. 
One way to mitigate effects of noise is to design algorithms with low circuit depths, but this is often challenging.
While some approaches help for specific applications~\cite{farhi2014quantum,peruzzo2014variational,mcclean2016theory,arguello2019analogue} or for individual circuits~\cite{peng2019simulating}, there are few general techniques for (re)designing low-depth quantum algorithms.
One technique is to trade shorter circuit depth for increased circuit width~\cite{broadbent2009parallelizing,abdessaied2013reducing,subasi2018entanglement}, i.e. use more qubits, but those qubits may still be unavailable or unacceptably noisy. 
New strategies are needed for designing low-width noise-resilient, NISQ algorithms.

One under-explored tool is qubit resetting, by which we mean the ability to reinitialize subsets of qubits in a known state, usually the $\ket{0}$ state, in the course of the computation~\cite{eggerWerninghaus2018reset}.
Generally, qubits are reinitialized either to prepare the entire apparatus to run a circuit or to reuse subsets of qubits in the course of a computation; we focus on the second use.
There exist methods for actively resetting qubits to their $\ket{0}$ state in time comparable to that required for a measurement~\cite{reedJohnson2010reset,geerlingsLeghtas2013reset,mcclurePaikGambetta2016reset,magnardKurpiers2018reset} --- with measurement and reset generally distinct processes.
The implementation depends on the particular hardware, but we are interested in resets as an algorithmic/software tool. 
This ability will be critical for error-correcting codes which require frequent stabilizing measurements~\cite{riste2019real}, and it has recently been used to design algorithms with reduced circuit width~\cite{huggins2019towards,LiuZhang2019variational,fossfeig2020holographic,rattew2020quantum}.
Work on automatically inserting resets as part of optimization and compiling should be particularly valuable for the first goal. Here, we contribute to the latter goal. 
We present algorithms for the application of entanglement spectroscopy which exploit qubit resets to achieve low circuit width while remaining noise-resilient.

Entanglement spectroscopy is the task of learning about the entanglement of a quantum state. 
The bipartite entanglement of a pure quantum state $\ket{\psi}$ on systems $A$ and $B$ can be characterized by the eigenvalues of the density operator of the reduced state $\rho_A = \tr_B(\ketbra{\psi}{\psi})$ (equivalent to the eigenvalues of $\rho_B$)~\cite{amico2008entanglement}. 
As noted by Li and Haldane, the entanglement spectrum (the eigenvalues of the so-called entanglement Hamiltonian $H$ defined via $\rho_A = e^{-H}$) contains much more information than the von Neumann entropy alone~\cite{li2008entanglement}.
For instance, it can be used to detect and characterize topological order and quantum phase transitions, as well as to determine whether a system obeys an area law and thus can be efficiently simulated classically~\cite{li2008entanglement,calabrese2008entanglement,fidkowski2010entanglement,prodan2010entanglement,de2012entanglement,swingle2012geometric,dalmonte2018quantum,johri2017entanglement,hermanns2017entanglement}.
Thus, Entanglement spectroscopy is an especially useful tool for analyzing outcomes of quantum simulation of many-body systems~\cite{islam2015measuring,linke2018measuring,kokail2020entanglement}. It may be similarly useful in characterizing the performance of NISQ devices.
Moreover, learning just the few largest eigenvalues of $\rho_A$, rather than performing full tomography, is often sufficient~\cite{johri2017entanglement}.
This task is computationally hard classically due to the exponentially growing dimension of the Hilbert space, making it a clear candidate for quantum algorithms.

Known efficient quantum algorithms to approximate the $n_{\text{max}}$ largest eigenvalues of $\rho_A$ generally begin by reducing the problem to computing
the traces of powers of the reduced density operator, i.e. $\tr(\rho_A^n)$ for $n=1,\dots,n_\text{max}$~\cite{song2012bipartite,johri2017entanglement}.\footnote{An exception to this is qPCA~\cite{lloyd2014quantum}, where the spectrum of a state $\rho$ is obtained using a different approach requiring phase estimation. This approach is not NISQ-friendly, so we do not discuss it here.}
These algorithms compute $\tr\left(\rho_A^n\right)$ using $O(n)$ copies of the state $\ket{\psi}$~\cite{johri2017entanglement,subasi2018entanglement}. The standard algorithm is an extension of the \SwapTest{}~\cite{buhrman2001quantum,gottesman2001quantum} by~\cite{johri2017entanglement} which we call the \HTlong{} (\HT{}) (Fig.~\ref{fig:JSTalg_HTestwithPermutation}). It uses $n$ copies of the state and has depth linear in $n$ and the size of the state (number of qubits).
The recent \TCTlong{} (\TCT{}) (Fig.~\ref{fig:tc_alg}) by~\cite{subasi2018entanglement} uses $2n$ copies of the state and achieves constant depth.
Although the latter algorithm achieves a depth suitable for NISQ devices, the linear width in $n$ of both algorithms will likely restrict their application to small $n$ in the NISQ era.

Having reduced entanglement spectroscopy to computing $\tr(\rho_A^n)$, we may state the task we study in this paper formally. \textbf{Problem:} Given as input a parameter $n$ and black-box access to a circuit preparing a pure state $\ket{\psi}$ on subsystems $A,B$, estimate $\tr(\rho_A^n)$.

In this work, we introduce new \emph{qubit-efficient} variants of the \HT{} and \TCT{} algorithms that require a number of qubits sufficient to prepare three or fewer copies of $\ket{\psi}$, independent of $n$. 
This is an asymptotically lower width than any previous efficient algorithm for computing $\tr(\rho_A^n)$ or the R\'{e}nyi entropy of order $n$.\footnote{The algorithm in \cite{subramanian2019quantum} also uses a number of qubits independent of $n$, and in contrast to the polynomial-time algorithms described in this work, it can be used to compute $\tr(\rho_A^\alpha)$ for non-integer $\alpha$. However, its time complexity scales exponentially in the system size.
} 
We achieve this by using qubit resets and preparing additional copies of the state in previously used registers, allowing us to perform computations on many copies of the state while using few qubits.

The depths of our qubit-efficient algorithms are linear in $n$ and the size of the state, but, crucially, our new algorithms do not suffer as much in the presence of noise as their increased depth suggests. 
Intuitively, one hopes that periodically resetting qubits prevents errors from accumulating, but because the resets only affect a subset of the qubits at a time, errors might still carry over.
By carefully choreographing the resets in our new algorithms, we try to prevent this from happening as much as possible.

We test our algorithms numerically and find that our qubit-efficient algorithms perform nearly identically well in the presence of noise as their higher-width analogs. We also implement one of our \qeTCTlong{} algorithms experimentally on the Honeywell System Model H0~\cite{pino2020demonstration}, estimating $\tr(\rho^n)$ for larger $n$ than possible on the device using previous algorithms.

Motivated by our results, we propose a generalization of circuit depth, which we call \emph{effective circuit depth}, for predicting the performance of quantum algorithms that use qubit resets on noisy devices. This new attribute helps explain why our qubit-efficient algorithms perform comparably to their original counterparts; for example, while the depths of our qubit-efficient \TCT{} variants are asymptotically greater than that of the original \TCT{}, their effective depths match up to a constant factor. Effective circuit depth is a better descriptor of quantum circuits with qubit resets than standard circuit depth and should aid in analyzing and designing future qubit-efficient algorithms.

\begin{table}[t]
\tabcolsep=0.04cm
\begin{small}
\begin{tabular}{|l|l|l|l|l|}
\hline
              & Width  & Depth                       & Effective depth             & Ref.                          \\ \hline
\HT{}         & $kn+1$ & $\tsp{}+ O(kn)$                     & $O(kn)$                     & \cite{johri2017entanglement}  \\
\qeHTfourk{}  & $4k+1$ & $\Theta(n\times(\tsp{}+k))$ & $\Theta(n\times(\tsp{}+k))$ & *                             \\
\qeHTthreek{} & $3k+1$ & $\Theta(n\times(\tsp{}+k))$ & $\Theta(n\times(\tsp{}+k))$ & *                             \\
\TCT{}        & $4kn$  & $\tsp{}+O(1)$               & $\tsp{}+O(1)$               & \cite{subasi2018entanglement} \\
\qeTCTsixk{}  & $6k$   & $\Theta(n\times(\tsp{}+1))$ & $\tsp{}+O(1)$               & *                             \\
\qeTCTfourk{} & $4k$   & $\Theta(n\times(\tsp{}+1))$ & $\tsp{}+O(1)$               & *                             \\ \hline
\end{tabular}
\end{small}
\caption{A summary of the algorithms in this paper for computing $\tr(\rho_A^n)$ of a state $\ket{\psi}_{AB}$ on $2k$ qubits. Algorithms marked with `*' in the final column are new. See Sections~\ref{sec:previous_work} and~\ref{sec:qeff_alg} for more details.~\label{table:table1}}
\end{table}

We begin by reviewing previous algorithms in Section~\ref{sec:previous_work}. In Section~\ref{sec:qeff_alg}, we introduce our qubit-efficient variants. 
These algorithms are summarized in Table~\ref{table:table1}.
In Section~\ref{sec:numerics}, we present numerical simulations comparing the performance of the new and original algorithms in the presence of noise.
In Section~\ref{sec:experiment}, we report the results of an experimental implementation of the \qeTCTlong{} on the Honeywell System Model H0.
We introduce effective circuit depth in Section~\ref{sec:effecDepth}, followed by discussion in Section~\ref{sec:discuss}.
Details of the numerical simulations are in the Appendix.

\section{Previous Work}
\label{sec:previous_work}

Given an entangled pure state $\ket{\psi}$ defined on subsystems $A$ and $B$, discarding the qubits associated with subsystem $B$ produces a mixed state $\rho_A$. 
If subsystem $A$ of interest has $k$ qubits, a subsystem $B$ of equal size is sufficient to create any mixed state on $A$ (the converse of purification). In what follows, we will assume that registers $A$ and $B$ each have $k$ qubits --- so $\ket{\psi}$ is $2k$ qubits and $\rho_A$ is $k$ qubits.
It is straightforward to generalize the algorithms discussed in this paper to cases where the registers are different sizes, in part because $A$ subsystems only ever interact with other $A$ subsystems and $B$ subsystems with other $B$ subsystems.

Standard methods for entanglement spectroscopy begin
with the observation from~\cite{song2012bipartite,johri2017entanglement} that traces of powers of the reduced density operator, i.e. $\tr(\rho_A^n)$ for $n=1,\dots,n_\text{max}$, can be used to approximately reconstruct the largest $n_\text{max}$ eigenvalues of $\rho_A$ via the Newton-Girard method.
This is especially useful given we are often interested in a small number $n_{\text{max}}\ll 2^k$ of the largest eigenvalues.
Alternatively, $\tr(\rho_A^n)$ can be used to exactly compute the R\'{e}nyi entropy of order $n$ (see, e.g.~\cite{subasi2018entanglement}).

Such traces can be expressed as the expectation values of the unitary cyclic permutation operators $P^\text{cyc}_A$ over $n$ copies of the state $\rho=\ketbra{\psi}{\psi}$,\footnote{See, e.g. Theorem 4.13 and Definition 3.15 of~\cite{buadescu2019quantum} for a proof.} 
where the subscript indicates that it acts only on the $A$ subsystems of the copies of $\rho$, i.e.
\begin{align}
    \label{eq:obs}
    \tr\left( \rho_A^n\right) = \langle P^\text{cyc}_{A} \rangle_{\rho_A^{\otimes n}} = \langle P^\text{cyc}_{A} \rangle_{\rho^{\otimes n}} .
\end{align}
Requiring $n$ copies is optimal for computing an $n$-th degree polynomial of $\rho$ since unitary evolution is linear~\cite{brun2004measuring}.
Using the permutation operator is a generalization of the premise for the well-known \SwapTest{}~\cite{buhrman2001quantum,gottesman2001quantum}, where the \SWAP{} gate is a cyclic permutation over two qubits.
Together, the previous two paragraphs reduce the problem of entanglement spectroscopy to estimating the expectation value of a unitary operator.
\begin{figure}[h]
    \centering%
    \hspace{0.5\columnsep}%
    \begin{minipage}{0.4\mylength}
        \begin{equation*}
            \Qcircuit @C=0.4em @R=1em {
            \lstick{\ket{0}} & \gate{H} & \ctrl{1} & \gate{H} & \meter \\
            \lstick{\ket{\Psi}} & \slash\qw & \gate{M} & \qw
            }
        \end{equation*}
    \end{minipage}
    \hspace{0.09\columnwidth}
    \begin{minipage}{0.45\mylength}%
        \begin{equation*}
            \Qcircuit @C=0.4em @R=1em {
            \lstick{\ket{\Psi}} & \slash\qw & \qw & \qw & \multigate{1}{\parbox{6em}{Bell basis\\measurement}} \\
            \lstick{\ket{\Psi}} & \slash\qw & \qw & \gate{M}  & \ghost{\parbox{6em}{Bell basis\\measurement}}
            }
        \end{equation*}
    \end{minipage}%
    \caption{(a) The Hadamard Test is on the left. It applies controlled-$M$ and requires one copy of $\ket{\Psi}$ and one ancilla qubit. The expectation values of the Pauli $Z$ and $-Y$ operators of the ancilla qubit are the real and imaginary parts of $\bra{\Psi}M\ket{\Psi}$, respectively.
    (b) The Two-Copy Test is on the right. It requires two copies of $\ket{\Psi}$ and applies $M$ to one of them. Performing an overlap measurement using the Bell basis algorithm gives 
    $\abs{\bra{\Psi}M\ket{\Psi}}^2$.}
    \label{fig:hadamard+twocopy}
\end{figure}
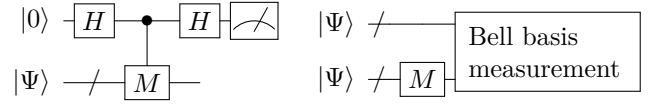

\begin{figure}[h]
    \centering
    \hspace{0.5\columnsep}%
    \begin{minipage}{0.43\mylength}
        \begin{equation*}
        \Qcircuit @C=0.4em @R=0.0em @!R {
    		\lstick{\ket{0}} 	& \gate{H} 	& \ctrl{1} 		& \ctrl{1} 		& \qw 	& \ctrl{1}		& \ctrl{1} 		& \ctrl{1}		& \gate{H} & \meter \\
    		\lstick{\rho_A} 	& {/} \qw 	& \qswap        & \qswap        & \qw 	& \qswap		& \qswap 		& \qswap		& \qw \\
    		\lstick{\rho_A} 	& {/} \qw 	& \qw\qwx		& \qw\qwx       & \qw 	& \qw\qwx 		& \qw\qwx		& \qswap\qwx	& \qw \\
    		\lstick{\rho_A} 	& {/} \qw 	& \qw\qwx       & \qw\qwx		& \qw 	& \qw\qwx		& \qswap\qwx	& \qw			& \qw \\
    		\lstick{\rho_A} 	& {/} \qw 	& \qw\qwx       & \qw\qwx     	& \qw 	& \qswap\qwx	& \qw			& \qw			& \qw \\
    		\cdots 				&			& \qwx			& \qwx			&		&														\\
    		\lstick{\rho_A} 	& {/} \qw 	& \qw\qwx       & \qswap\qwx    & \qw 	& \qw		 	& \qw			& \qw			& \qw  \\
    		\lstick{\rho_A} 	& {/} \qw 	& \qswap\qwx    & \qw           & \qw 	& \qw			& \qw			& \qw			& \qw
	    }
    	\end{equation*}
    \end{minipage}%
    \hspace{0.1\mylength}%
    \begin{minipage}{0.43\mylength}%
        \begin{equation*}
        \Qcircuit @C=0.4em @R=0.0em @!R {
    		\lstick{\ket{0}} 	& \gate{H} 	& \ctrl{1} 		& \ctrl{1} 		& \ctrl{1}  	& \qw 	& \ctrl{1}		& \ctrl{1} 	& \gate{H} & \meter \\
    		\lstick{\rho_A} 	& {/} \qw 	& \qswap        & \qswap        & \qswap		& \qw 	& \qswap		& \qswap 		& \qw \\
    		\lstick{\rho_A} 	& {/} \qw 	& \qswap\qwx	& \qw\qwx       & \qw\qwx		& \qw 	& \qw\qwx 		& \qw\qwx		& \qw \\
    		\lstick{\rho_A} 	& {/} \qw 	& \qw           & \qswap\qwx	& \qw\qwx 		& \qw 	& \qw\qwx		& \qw\qwx		& \qw \\
    		\lstick{\rho_A} 	& {/} \qw 	& \qw           & \qw           & \qswap\qwx 	& \qw 	& \qw\qwx 		& \qw\qwx		& \qw \\
    		\cdots 				&			&				&				&				&		& \qwx			& \qwx		\\
    		\lstick{\rho_A} 	& {/} \qw 	& \qw           & \qw           & \qw 			& \qw 	& \qswap\qwx 	& \qw\qwx		& \qw  \\
    		\lstick{\rho_A} 	& {/} \qw 	& \qw           & \qw           & \qw 			& \qw 	& \qw			& \qswap\qwx	& \qw
	    }
	    \end{equation*}
    \end{minipage}
    \caption{The \HT{} algorithm, which is the \HadamardTest{} with a cyclic permutation operator, computes $\tr\left(\rho_A^n\right)$. 
    As the two circuits show, $P^\text{cyc}_{A}$ can be implemented as either a left shift or right shift, respectively. 
    Each \CSWAP{} shown is implicitly implemented by $k$ sequential \CSWAP{}s.
    The circuit depth is $\tsp{}+\Theta\left(kn\right)$ and the width is $2kn+1$.
    }
	\label{fig:JSTalg_HTestwithPermutation}
\end{figure}
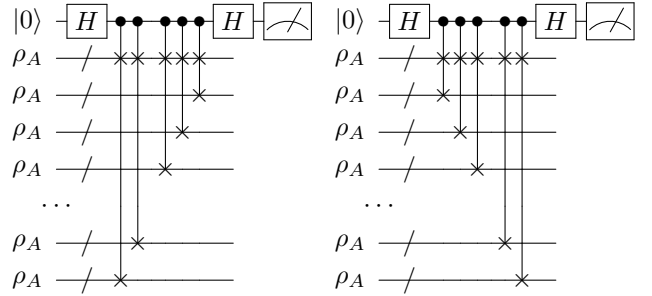

Note that there are several definitions and implementations of the cyclic permutation operator that are equivalent for the purposes of entanglement spectroscopy. 
First, a cyclic permutation may be a left shift or a right shift, shifting the contents of the first register back to the $n$-th register or forwards to the second register, respectively. These different definitions are illustrated in Fig.~\ref{fig:JSTalg_HTestwithPermutation}.
Second, either type of shift can be implemented using $n-1$ transpositions (swaps), but there are many possible decompositions. For example, using cycle notation, $\left(4 1 2 3\right)$ is equivalent to $\left(3 4\right) \left(2 3\right) \left(1 2\right)$ and to $\left(1 2\right) \left(1 3\right) \left(1 4\right)$.
Eq.~\ref{eq:obs} holds for all of these variants.
Our choices of when to use left shift and right shift and our choice of decomposition are arbitrary.

\subsection{\HTlong{} (\HT{})}
\label{sec:HT_alg}

The standard algorithm for estimating the expectation value of an arbitrary unitary operator $M$ on a state $\ket{\Psi}$ is the \HadamardTest{}, illustrated and described in Fig.~\ref{fig:hadamard+twocopy}(a). To be clear, the real part of $\bra{\Psi}M\ket{\Psi}$ is calculated by $p_0-p_1$, where $p_i$ denotes the probability that the ancilla qubit is in state $\ket{i}$. 
The estimates of $p_0,p_1$ converge with accuracy $O(1/\sqrt{S})$ where $S$ is the number of times the circuit is run
(this scaling can be improved using quantum amplitude estimation~\cite{johri2017entanglement}, but this is impractical in the NISQ era).

For entanglement spectroscopy, we substitute the cyclic permutation operator $P^\text{cyc}_A$ for $M$ and $\ket{\psi}^{\otimes n}$ for $\ket{\Psi}$. 
This is the algorithm of Johri, Steiger, and Troyer~\cite{johri2017entanglement}, which we refer to as the \HTlong (\HT).\footnote{In \cite{subasi2018entanglement}, this was referred to as the JST algorithm after the initials of the authors of \cite{johri2017entanglement}. Here, we opt to use a more descriptive name which is easier to extend to our new algorithms.}
It is illustrated in Fig.~\ref{fig:JSTalg_HTestwithPermutation}.
\HT{} is a generalization of the \SwapTest{}~\cite{buhrman2001quantum,gottesman2001quantum}, where the \SwapTest{} is the \HadamardTest{} with $M=\SWAP{}$ acting on two states $\rho,\sigma$ to compute the overlap $\tr\left(\SWAP{} \rho \otimes \sigma\right) =\tr\left(\rho\sigma\right)$, which equals the purity $\tr\left(\rho^2\right)$ when $\rho=\sigma$.

A swap of two $k$-qubit registers can be implemented with $k$ swaps of individual qubits, so the total number of controlled-\SWAP{} (\CSWAP{}) gates in \HT{} is $k(n-1)$. 
A downside of \HT{} is that all of the swap operations are controlled on the same ancilla, so the \CSWAP{} gates must be applied sequentially. Given that \CSWAP{} has constant depth, the circuit depth of \HT{} scales linearly with $k n$. 
Specifically, it is $\tsp{}+\tcs kn+2\thad  = \tsp{}+\Theta\left(kn\right)$, where $\tsp{}$ is the time it takes to prepare a single copy of the state, $\tcs{}$ is the time to implement a \CSWAP{}, and $\thad{}$ is the time to implement a Hadamard gate.
Note that we treat $\tcs{}$ and $\thad{}$ as constants which depend on the hardware and decompositions used, independent of the input; we also assume all-to-all connectivity.
The circuit width of \HT{} is $2 k n+1=\Theta(kn)$, including the qubits for the $B$ subsystems --- recall $\ket{\psi}$ is $2k$ qubits and each subsystem is $k$ qubits.

We just stated the depth of \HT{} as $\tsp+\Theta(kn)$, but previous work has usually stated the depth as being linear in $kn$, dropping the time for state preparation.
In the model considered in previous work, 
algorithms accept many copies of $\ket{\psi}$ as input at the beginning, as in Fig.~\ref{fig:hadamard+twocopy}. Thus, state preparation and the algorithm are considered independently.
Our new algorithms will require a setting in which state preparation is intertwined with the rest of the algorithm.
This is also a reasonable setting since the states used as input in the previous setting are prepared by some physical procedure which we may represent by a subcircuit as in Fig.~\ref{fig:H_qubiteff_4k}.
Rather than many copies of $\ket{\psi}$, we assume that a description of the state preparation circuit is given as input. (In fact, our algorithms work in the more restricted black box model, in that we only consider the output of the state preparation circuit, never examining the circuit itself.)
To fairly compare the new and previous algorithms, we assume the same setting, so we include state preparation in the depths of all algorithms.

Finally, we note that in the setting where state preparation is included as part of the algorithm, a simple modification can improve noise resilience. Instead of preparing all copies at the start of the algorithm, state preparation should be delayed until needed. For example, in the second \HT{} circuit in Fig.~\ref{fig:JSTalg_HTestwithPermutation}, the preparation of the third copy could be delayed by $\tcs{}$ compared to the preparation of the first copy.

\subsection{\TCTlong (\TCT)}
\label{sec:TCT_alg}
Recently, Cincio, Suba\c{s}\i, Sornborger, and Coles~\cite{cincio2018learning} rediscovered an algorithm of Garcia-Escartin and Chamorro-Posada~\cite{garcia2013BBA} for computing the overlap between two states using Bell basis measurements of corresponding pairs of qubits from each state followed by efficient classical post-processing.
For intuition, the Bell basis is an eigenbasis of the \SWAP{} operator, allowing a Bell basis measurement to reproduce the result of the \SwapTest{}.
Ref.~\cite{garcia2013BBA} related this algorithm to the Hong-Ou-Mandel effect in the context of quantum optics and referred to it as a destructive \SwapTest{}, while~\cite{cincio2018learning} emphasized that it can be implemented with constant depth in a quantum computer. 
We refer to this algorithm as the Bell basis algorithm.

\begin{figure*}[]
    \begin{minipage}{0.49\linewidth}
        \begin{equation*}
        \Qcircuit @C=0.8em @R=0.0em @!R {
    		\lstick{} & \slash\qw & \qw & \qw 			& \ctrl{7}	& \qw 		& \qw 		& \qw 		& \qw 		& \qw		& \gate{H} & \meter \\
    		\lstick{} & \slash\qw & \qw & \qw 			& \qw		& \ctrl{7} 	& \qw 		& \qw 		& \qw		& \qw		& \gate{H} & \meter
    			\inputgroupv{1}{2}{1em}{1em}{\ket{\psi}} \\
    		\lstick{} & \slash\qw & \qw & \qw 			& \qw 		& \qw 		& \ctrl{7} 	& \qw 		& \qw		& \qw		& \gate{H} & \meter \\
    		\lstick{} & \slash\qw & \qw & \qw 			& \qw 		& \qw 		& \qw 		& \ctrl{7} 	& \qw		& \qw		& \gate{H} & \meter
    			\inputgroupv{3}{4}{1em}{1em}{\ket{\psi}} \\
    		\lstick{} & \slash\qw & \qw & \qw 			& \qw 		& \qw 		& \qw	 	& \qw 		& \ctrl{7} 	& \qw		& \gate{H} & \meter \\
    		\lstick{} & \slash\qw & \qw & \qw 			& \qw 		& \qw 		& \qw 		& \qw	 	& \qw		& \ctrl{7}	& \gate{H} & \meter
    			\inputgroupv{5}{6}{1em}{1em}{\ket{\psi}} \\
    				  &				&	  &					&			&			&			&			&			&			&						&		\\
    		\lstick{} & \slash\qw & \qw & \link{4}{-1}	& \targ 	& \qw 		& \qw	 	& \qw 		& \qw		& \qw		& \qw					& \meter \\
    		\lstick{} & \slash\qw & \qw & \qw				& \qw 		& \targ 	& \qw 		& \qw 		& \qw		& \qw 		& \qw					& \meter
    			\inputgroupv{8}{9}{1em}{1em}{\ket{\psi}} \\
    		\lstick{} & \slash\qw & \qw & \link{-2}{-1}	& \qw 		& \qw 		& \targ		& \qw 		& \qw 		& \qw		& \qw					& \meter \\
    		\lstick{} & \slash\qw & \qw & \qw				& \qw 		& \qw 		& \qw 		& \targ 	& \qw 		& \qw		& \qw					& \meter
    			\inputgroupv{10}{11}{1em}{1em}{\ket{\psi}} \\
    		\lstick{} & \slash\qw & \qw & \link{-2}{-1}	& \qw 		& \qw 		& \qw 		& \qw		& \targ		& \qw		& \qw					& \meter \\
    		\lstick{} & \slash\qw & \qw & \qw				& \qw 		& \qw 		& \qw 		& \qw	 	& \qw 		& \targ		& \qw					& \meter
    			\inputgroupv{12}{13}{1em}{1em}{\ket{\psi}}
    		\gategroup{7}{1}{7}{12}{0em}{--} %
    	}
	    \end{equation*}
    \end{minipage}
    \begin{minipage}{0.49\linewidth}%
        \begin{equation*}
        \Qcircuit @C=0.8em @R=0.0em @!R {
    		\lstick{} & \slash\qw & \qw & \qw 			& \ctrl{11}	& \qw 		& \qw 		& \qw 		& \qw 		& \qw		& \gate{H} & \meter \\
    		\lstick{} & \slash\qw & \qw & \qw 			& \qw		& \ctrl{7} 	& \qw 		& \qw 		& \qw		& \qw		& \gate{H} & \meter
    			\inputgroupv{1}{2}{1em}{1em}{\ket{\psi}} \\
    		\lstick{} & \slash\qw & \qw & \qw 			& \qw 		& \qw 		& \ctrl{5} 	& \qw 		& \qw		& \qw		& \gate{H} & \meter \\
    		\lstick{} & \slash\qw & \qw & \qw 			& \qw 		& \qw 		& \qw 		& \ctrl{7} 	& \qw		& \qw		& \gate{H} & \meter
    			\inputgroupv{3}{4}{1em}{1em}{\ket{\psi}} \\
    		\lstick{} & \slash\qw & \qw & \qw 			& \qw 		& \qw 		& \qw	 	& \qw 		& \ctrl{5} 	& \qw		& \gate{H} & \meter \\
    		\lstick{} & \slash\qw & \qw & \qw 			& \qw 		& \qw 		& \qw 		& \qw	 	& \qw		& \ctrl{7}	& \gate{H} & \meter
    			\inputgroupv{5}{6}{1em}{1em}{\ket{\psi}} \\
    				  &				&	  &					&			&			&			&			&			&			&						&		\\
    		\lstick{} & \slash\qw & \qw & \qw				& \qw	 	& \qw 		& \targ	 	& \qw 		& \qw		& \qw		& \qw					& \meter \\
    		\lstick{} & \slash\qw & \qw & \qw				& \qw 		& \targ 	& \qw 		& \qw 		& \qw		& \qw 		& \qw					& \meter
    			\inputgroupv{8}{9}{1em}{1em}{\ket{\psi}} \\
    		\lstick{} & \slash\qw & \qw & \qw				& \qw 		& \qw 		& \qw		& \qw 		& \targ		& \qw		& \qw					& \meter \\
    		\lstick{} & \slash\qw & \qw & \qw				& \qw 		& \qw 		& \qw 		& \targ 	& \qw 		& \qw		& \qw					& \meter
    			\inputgroupv{10}{11}{1em}{1em}{\ket{\psi}} \\
    		\lstick{} & \slash\qw & \qw & \qw				& \targ		& \qw 		& \qw 		& \qw		& \qw		& \qw		& \qw					& \meter \\
    		\lstick{} & \slash\qw & \qw & \qw				& \qw 		& \qw 		& \qw 		& \qw	 	& \qw 		& \targ		& \qw					& \meter
    			\inputgroupv{12}{13}{1em}{1em}{\ket{\psi}}
    		\gategroup{7}{1}{7}{12}{0em}{--} %
	    }
        \end{equation*}
    \end{minipage}
    \caption{(a) On the left is the \TwoCopyTest{} with a $P^\text{cyc}_A$ gate for computing $\tr\left(\rho_A^n\right)^2$ for $n=3$. Gates on registers are implicitly applied to all qubits in parallel: for example, $H$ is $H^{\otimes k}$. The top and bottom wires for each state contain subsystems $A$ and $B$, respectively. The classical post-processing step is as in Eq.~\eqref{eq:bba_postprocessing}. (b) On the right is the \TCT{} algorithm, developed from the circuit on the left by indirectly implementing the cyclic permutation by reindexing the \CNOT{} gates. The classical post-processing is as in Eq.~\eqref{eq:tct-postprocessing}.
    The circuit depth is $\tsp{}+O(1)$ and the width is $4kn$.}
	\label{fig:tc_alg}
\end{figure*}
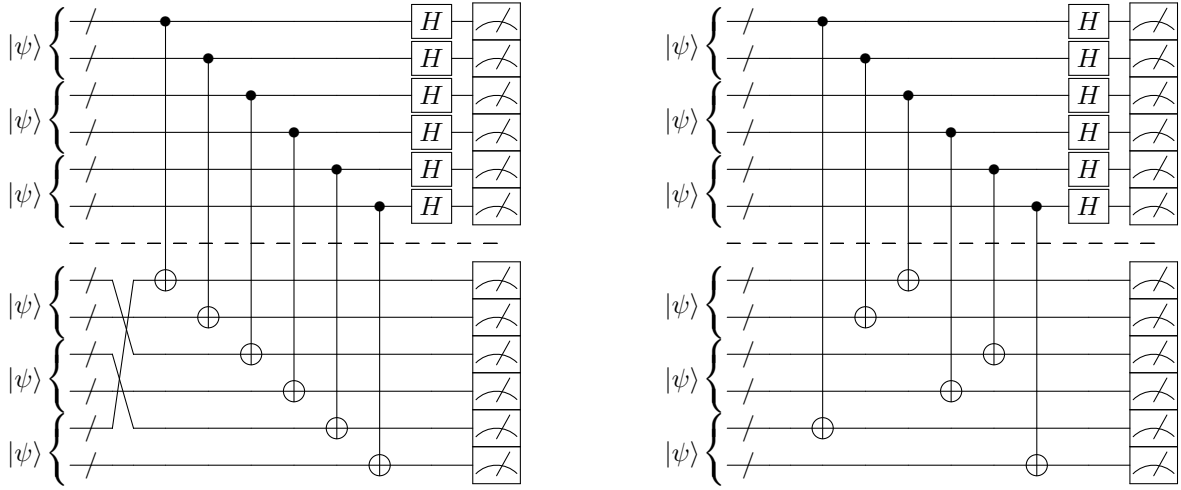

For completeness, a Bell basis measurement on a pair of qubits involves applying a controlled-not (\CNOT{}) gate and then a Hadamard on one of the qubits followed by measuring in the standard basis; see Fig.~\ref{fig:tc_alg}(a) for an example.
Importantly, because each \CNOT{} acts on a different pair of qubits, the measurement can be performed with a single layer of \CNOT{}s and a single layer of Hadamards, which is constant depth.
To then compute the overlap between two states $\rho$ and $\sigma$ each of size $m$, the classical post-processing step is to compute the linear function 
\begin{equation}\label{eq:bba_postprocessing}
    \sum_{r,s\in\{0,1\}^m} (-1)^{r_1s_1+\dots + r_ms_m} ~p_{r,s} ~,
\end{equation} 
where $p_{r,s}$ is the experimentally measured frequency that the first $m$ qubits, corresponding to $\rho$, are measured in state $\ket{r}$ and that the second $m$ qubits, corresponding to $\sigma$, are measured in state $\ket{s}$. This classical step can be performed in time linear in the number of trials. 

Building on the Bell basis algorithm, Suba\c{s}\i, Cinicio, and Coles~\cite{subasi2018entanglement} introduced an algorithm for estimating $\abs{\bra{\Psi} M \ket{\Psi}}^2$ for a unitary operator $M$, which is called the \TwoCopyTest{}. Classical post-processing can then yield the magnitude of the expected value, $\abs{\bra{\Psi} M \ket{\Psi}}=\abs{\tr(M\ketbra{\Psi}{\Psi})}$.
This algorithm relies on the observation that for pure states, the squared expectation value $\abs{\bra{\Psi} M \ket{\Psi}}^2$ is equivalent to the overlap between states $\ket{\Psi}$ and $M\ket{\Psi}$. 
As depicted in Fig.~\ref{fig:hadamard+twocopy}(b), the \TwoCopyTest{} accepts two copies of the state $\ket{\Psi}$, applies $M$ to one, and performs an overlap measurement using the Bell basis algorithm.
This requires enough qubits for two copies of the state. 
Because the Bell basis measurement is constant-depth, the depth of the overall algorithm only depends on $M$ and \tsp{}.
Unlike the \HadamardTest{}, the \TwoCopyTest{} cannot be used to obtain the real and imaginary parts of $\bra{\Psi} M \ket{\Psi}$. 
Also, while the \HadamardTest{} works both for pure states and for mixed states, the \TwoCopyTest{} can only be used to compute expectation values for pure states.
Estimating $\abs{\bra{\Psi} M \ket{\Psi}}$ using the \TwoCopyTest{} converges with accuracy $O(1/\sqrt{S})$, where $S$ is the number of times the circuit is run.

A crucial difference between the \HadamardTest{} and the \TwoCopyTest{} is that the latter uses the unitary $M$ instead of controlled-$M$. 
Recalling that \HT{} has linear depth because the controlled gates have to be applied sequentially, eliminating the control not only reduces the gate count, it also allows for the possibility of parallelization. Given the right operator $M$, this can lead to applications of the \TwoCopyTest{} with very shallow circuits.

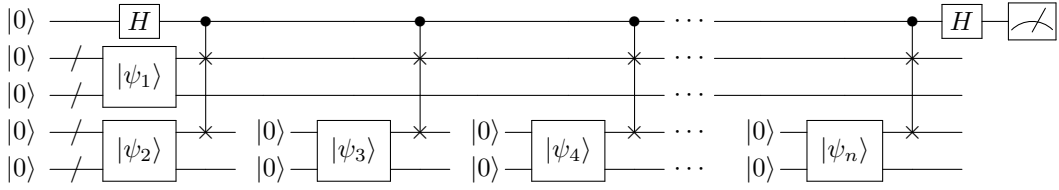
\begin{figure*}[t]
    \begin{equation*}
    \Qcircuit @C=1.0em @R=0.0em @!R {
    	\lstick{\ket{0}} & \qw			& \gate{H} 							& \ctrl{1} 		& \qw 												& \qw 							& \ctrl{1} 		& \qw											 	& \qw 								& \ctrl{1} 	 & \push{\rule{.2em}{0em}\cdots\rule{.2em}{0em}}\qw & \qw											 	& \qw 								& \ctrl{1} 		& \gate{H} & \meter \\
    	\lstick{\ket{0}} & \slash\qw	& \multigate{1}{\ket{\psi_{1}}} 	& \qswap 		& \qw 												& \qw 							& \qswap 		& \qw											 	& \qw 								& \qswap 	 & \push{\rule{.2em}{0em}\cdots\rule{.2em}{0em}}\qw & \qw											 	& \qw 								& \qswap		& \qw	\\
    	\lstick{\ket{0}} & \slash\qw	& \ghost{\ket{\psi_{1}}} 			& \qw\qwx 		& \qw 												& \qw 							& \qwx\qw 		& \qw											 	& \qw								& \qw\qwx	 & \push{\rule{.2em}{0em}\cdots\rule{.2em}{0em}}\qw & \qw											 	& \qw								& \qw\qwx		& \qw 	\\
    	\lstick{\ket{0}} & \slash\qw	& \multigate{1}{\ket{\psi_{2}}} 	& \qswap\qwx 	& \push{\rule{.6em}{0em}\ket{0}\rule{0em}{0em}}\qw 	& \multigate{1}{\ket{\psi_{3}}} & \qswap\qwx 	& \push{\rule{.6em}{0em}\ket{0}\rule{0em}{0em}}\qw 	& \multigate{1}{\ket{\psi_{4}}} 	& \qswap\qwx & \push{\rule{.2em}{0em}\cdots\rule{.2em}{0em}}\qw & \push{\rule{0em}{0em}\ket{0}\rule{0em}{0em}} 	& \multigate{1}{\ket{\psi_{n}}} 	& \qswap\qwx	& \qw 	\\
    	\lstick{\ket{0}} & \slash\qw	& \ghost{\ket{\psi_{2}}} 			& \qw			& \push{\rule{.6em}{0em}\ket{0}\rule{0em}{0em}}\qw 	& \ghost{\ket{\psi_{3}}} 		& \qw			& \push{\rule{.6em}{0em}\ket{0}\rule{0em}{0em}}\qw 	& \ghost{\ket{\psi_{4}}} 			& \qw		 & \push{\rule{.2em}{0em}\cdots\rule{.2em}{0em}}\qw & \push{\rule{0em}{0em}\ket{0}\rule{0em}{0em}} 	& \ghost{\ket{\psi_{n}}} 			& \qw			& \qw
    }
    \end{equation*}
    \caption{The \qeHTfourklong{}. A break in a wire followed by a new $\ket{0}$ indicates a reset. Each $\ket{\psi_i}$ indicates the preparation of another copy of $\ket{\psi}$; the subscripts are only for guidance. The circuit depth and effective depth are $\Theta(n\times(\tsp{}+k))$ and the circuit width is $4k+1$.}
    \label{fig:H_qubiteff_4k}
\end{figure*}
\begin{figure*}[t]
    \begin{equation*}
    \Qcircuit @C=1.0em @R=0.0em @!R {
        \lstick{\ket{0}} & \qw			& \gate{H} 							& \qw												& \qw 							& \ctrl{1} 		& \qw 												& \qw 							& \ctrl{1} 		& \push{\rule{.2em}{0em}\cdots\rule{.2em}{0em}}\qw & \qw 											& \qw 							& \ctrl{1}		& \gate{H} & \meter \\
        \lstick{\ket{0}} & \slash\qw 	& \multigate{1}{\ket{\psi_{1}}} 	& \qw												& \qw							& \qswap 		& \qw 												& \qw 							& \qswap 		& \push{\rule{.2em}{0em}\cdots\rule{.2em}{0em}}\qw & \qw 											& \qw 							& \qswap 		& \qw	\\
        \lstick{\ket{0}} & \slash\qw 	& \ghost{\ket{\psi_{1}}} 			& \push{\rule{.6em}{0em}\ket{0}\rule{0em}{0em}}\qw	& \multigate{1}{\ket{\psi_{2}}}	& \qswap\qwx 	& \push{\rule{.6em}{0em}\ket{0}\rule{0em}{0em}}\qw 	& \multigate{1}{\ket{\psi_{3}}} & \qswap\qwx	& \push{\rule{.2em}{0em}\cdots\rule{.2em}{0em}}\qw & \push{\rule{0em}{0em}\ket{0}\rule{0em}{0em}} 	& \multigate{1}{\ket{\psi_{n}}} & \qswap\qwx	& \qw	\\
        \lstick{\ket{0}} & \slash\qw 	& \qw								& \qw 												& \ghost{\ket{\psi_{2}}}	 	& \qw 			& \push{\rule{.6em}{0em}\ket{0}\rule{0em}{0em}}\qw 	& \ghost{\ket{\psi_{3}}} 		& \qw 			& \push{\rule{.2em}{0em}\cdots\rule{.2em}{0em}}\qw & \push{\rule{0em}{0em}\ket{0}\rule{0em}{0em}} 	& \ghost{\ket{\psi_{n}}} 		& \qw 			& \qw
    }
    \end{equation*}
    \caption{The \qeHTthreeklong{}. The circuit depth and effective depth are $\Theta(n\times(\tsp{}+k))$, asymptotically the same as the \qeHTfourk{}, and the circuit width is $3k+1$.}
    \label{fig:H_qubiteff_3k}
\end{figure*}
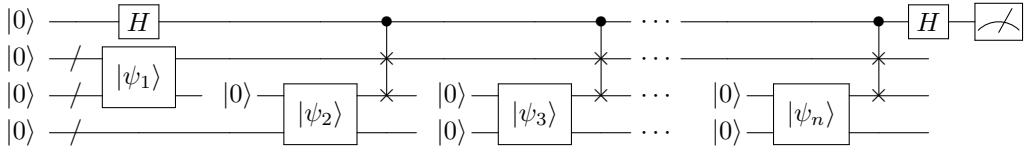

The \TwoCopyTest{} can be applied to entanglement spectroscopy by observing that Eq.~\eqref{eq:obs} can be recovered from
$\abs{\bra{\Psi} M \ket{\Psi}}^2 = \tr\left(\rho_A^n\right)^2$ with the choices $M = P^\text{cyc}_A$ and $\ket{\Psi} = \ket{\psi}^{\otimes n}$ combined with the fact $\tr\left(\rho_A^n\right)$ is real and nonnegative.
This circuit is depicted in Fig.~\ref{fig:tc_alg}(a).
Unlike \HT{}, this requires access to the full state $\ket{\psi}$. Next, since an (uncontrolled) permutation operator is equivalent to a relabeling of the registers on which it acts, the cyclic permutation can be implemented without any gates by carefully changing the registers which the \CNOT{}s in the Bell basis measurement act on and reindexing the classical post-processing formula. We refer to this algorithm as the \TCTlong{} (\TCT{})~\cite{subasi2018entanglement}. This circuit is depicted in Fig.~\ref{fig:tc_alg}(b).

To be clear, let $\ket{\psi_i}$ denote the $i$-th state in the first copy of $\ket{\Psi}$ and $\ket{\psi_i'}$ denote the $i$-th state in the second copy of $\ket{\Psi}$, where the operator $P^\text{cyc}_A$ is applied to the second copy. Then, the $B$ subsystem of $\ket{\psi_i}$ is paired with the $B$ subsystem of $\ket{\psi_i'}$, and the $A$ subsystem of $\ket{\psi_i}$ is paired with the $A$ subsystem of $\ket{\psi_{i-1}'}$ when the permutation is a right shift ($\ket{\psi_{i+1}'}$ when a left shift). The edge case of $\ket{\psi_1}$ ($\ket{\psi_n}$ when a left shift) is handled by performing indexing modulo $n$.
The post-processing calculation, derived from Eq.~\eqref{eq:bba_postprocessing}, is  more complicated but still efficient. Assuming the permutation is a right shift, the formula for $\tr\left(\rho_A^n\right)$ is the square root of
\begin{equation}\label{eq:tct-postprocessing}
    \sum_{\substack{\vec{A_j},\vec{B_j},\vec{A'_j},\vec{B'_j} \in \{0,1\}^{k}\\ j\in[n] }} (-1)^{\sum_{\ell=1}^{n} \vec{A_\ell}\cdot\vec{A'_{\ell-1}} + \vec{B_\ell}\cdot\vec{B'_\ell}} ~p_{\vec{A_1},\vec{B_1},\dots} ~,
\end{equation}
where $p_{\vec{A_1},\vec{B_1},\dots}$ is the experimentally measured frequency that for all $i\in[n]$, the qubits initially containing the $A$ and $B$ subsystems of $\ket{\psi_i}$ are measured in the states $\ket{\vec{A_i}}$ and $\ket{\vec{B_i}}$, respectively, and that the qubits initially containing $\ket{\psi'_i}$ are measured in the states $\ket{\vec{A'_i}}$ and $\ket{\vec{B'_i}}$, respectively.
The circuit depth of \TCT{} is $\tsp{} + \tcn{} + \thad{} = \tsp{}+O(1)$, independent of $k$ and $n$, where $\tcn{}$ is the time to implement a \CNOT{} gate. This is asymptotically better than \HT{}, but comes with the tradeoff that the circuit width is $4kn$, almost twice the width of \HT{}.

\section{Qubit-Efficient Algorithms}
\label{sec:qeff_alg}

In this work, we give variations of the \HT{} and \TCT{} algorithms which achieve asymptotically lower circuit width --- proportional to $k$ but independent of $n$ --- without significantly increasing the susceptibility to noise.
We refer to these as \emph{qubit-efficient} \HT{} and \TCT{} algorithms.
For both the \HT{} and \TCT{}, we give two variants where one achieves lower width than the other; we do this in part because the higher-width variants are easier to understand.
The high-level idea we rely on is to prepare only as many copies of the state $\ket{\psi}$ at a time as necessary. 
The structures of both \HT{} and \TCT{} are such that every time a new copy is needed to interact with existing copies, one of the existing copies is finished, with no gates left to act on it. So, the latter copy's qubits may be measured, reset, and used to prepare the new copy of the state (the measurement is optional depending on the particular algorithm: for example, some error mitigation methods might require measurement results). This allows us to run the \HT{} and \TCT{} algorithms with a circuit width independent of $n$. These algorithms rely on the ability to reset qubits in the course of a quantum computation.
\subsection{Qubit-Efficient \HT}
\label{sec:QEH}

Observe that every register in the \HT{} circuit (Fig.~\ref{fig:JSTalg_HTestwithPermutation}) except for the ancilla qubit interacts with just two other registers and the ancilla. 
The state of the ancilla qubit, and so the output of the algorithm, is not affected by discarding other registers, so they can be reset and recycled once the last gate on them has been applied.
At any time, we just need enough qubits to prepare two copies of the state and the ancilla qubit. So, by resetting qubits when we are done with their contents, we can implement \HT{} using a constant number of registers.
Note that measuring the qubits before resetting them is unnecessary unless one wants to perform postselection~\cite{linke2018measuring,subasi2018entanglement}.

Our first algorithm implementing this qubit-efficient strategy is given in Fig.~\ref{fig:H_qubiteff_4k}. Recalling that $\ket{\psi}$ is a state on $2k$ qubits, the circuit width is $4k+1$, independent of $n$. We refer to this algorithm as the \qeHTfourk{}.

The action of the algorithm can be verified by computing the reduced density matrix of the ancilla qubit after the $m$-th controlled-SWAP operation:
\begin{align}
    \rho_{anc} &= \frac{1}{2} I + \frac{1}{2} \tr\left(\rho_A^m\right) X\, .
\end{align}
Thus, after $n$ controlled-SWAP operations, a measurement in the $X$ basis yields $\tr\left(\rho_A^n\right)$, as desired.

\begin{figure*}[t]
    \begin{equation*}
    \Qcircuit @C=1.0em @R=0.0em @!R {
        \lstick{\ket{0}} & \slash\qw	& \multigate{1}{\ket{\psi'_{1}}} & \qw 		& \qw		& \qw 		& \qw 		& \qw 											 & \qw						& \targ 	& \qw 		& \meter & \push{\rule{.2em}{0em}\ket{0}\rule{.2em}{0em}}	& \multigate{1}{\ket{\psi'_{2}}} 		& \qw 		& \qw 		& \qw 	 & \qw 												& \qw 						& \push{\rule{.2em}{0em}\cdots\rule{.2em}{0em}}\qw  \\
        \lstick{\ket{0}} & \slash\qw	& \ghost{\ket{\psi'_{1}}} 		& \qw	 	& \targ		& \qw  		& \meter 	& \push{\rule{.2em}{0em}\ket{0}\rule{.2em}{0em}} & \multigate{1}{\ket{\psi_{2}}} 	& \ctrl{-1} & \gate{H} 	& \meter & \push{\rule{.2em}{0em}\ket{0}\rule{.2em}{0em}} 	& \ghost{\ket{\psi'_{2}}} 			& \targ 	& \qw 		& \meter & \push{\rule{.2em}{0em}\ket{0}\rule{.2em}{0em}} 	& \multigate{1}{\ket{\psi_{3}}} 	& \push{\rule{.2em}{0em}\cdots\rule{.2em}{0em}}\qw  \\
        \lstick{\ket{0}} & \slash\qw	& \multigate{1}{\ket{\psi_{1}}} 	& \ctrl{2} 	& \qw		& \gate{H} 	& \meter 	& \push{\rule{.2em}{0em}\ket{0}\rule{.2em}{0em}} & \ghost{\ket{\psi_{2}}} 			& \qw 		& \qw 		& \qw 	 & \qw 												& \qw 							& \ctrl{-1} & \gate{H} 	& \meter & \push{\rule{.2em}{0em}\ket{0}\rule{.2em}{0em}} 	& \ghost{\ket{\psi_{3}}} 			& \push{\rule{.2em}{0em}\cdots\rule{.2em}{0em}}\qw  \\
        \lstick{\ket{0}} & \slash\qw	& \ghost{\ket{\psi_{1}}}		 	& \qw		& \ctrl{-2} & \gate{H} 	& \meter 	& \push{\rule{.2em}{0em}\ket{0}\rule{.2em}{0em}} & \multigate{1}{\ket{\psi_{n}}} 	& \qw 		& \qw 		& \qw 	 & \qw 												& \qw 							& \ctrl{1} 	& \gate{H} 	& \meter & \push{\rule{.2em}{0em}\ket{0}\rule{.2em}{0em}} 	& \multigate{1}{\ket{\psi_{n-1}}} & \push{\rule{.2em}{0em}\cdots\rule{.2em}{0em}}\qw  \\
        \lstick{\ket{0}} & \slash\qw	& \multigate{1}{\ket{\psi'_{n}}} & \targ		& \qw	 	& \qw 		& \meter 	& \push{\rule{.2em}{0em}\ket{0}\rule{.2em}{0em}} & \ghost{\ket{\psi_{n}}} 			& \ctrl{1} 	& \gate{H} 	& \meter & \push{\rule{.2em}{0em}\ket{0}\rule{.2em}{0em}} 	& \multigate{1}{\ket{\psi'_{n-1}}} 	& \targ 	& \qw 		& \meter & \push{\rule{.2em}{0em}\ket{0}\rule{.2em}{0em}} 	& \ghost{\ket{\psi_{n-1}}} 		& \push{\rule{.2em}{0em}\cdots\rule{.2em}{0em}}\qw  \\
        \lstick{\ket{0}} & \slash\qw	& \ghost{\ket{\psi'_{n}}} 		& \qw 		& \qw 		& \qw 		& \qw 		& \qw 											 & \qw 						& \targ		& \qw		& \meter & \push{\rule{.2em}{0em}\ket{0}\rule{.2em}{0em}}	& \ghost{\ket{\psi'_{n-1}}} 		& \qw 		& \qw 		& \qw 	 & \qw 												& \qw 						& \push{\rule{.2em}{0em}\cdots\rule{.2em}{0em}}\qw 
    }
    \end{equation*}
    \caption{The \qeTCTsixklong{}. The circuit depth is $\Theta(n\times(\tsp+1))$, the effective depth is $2(\tsp+O(1)) = \Theta(\tsp+1)$, and the width is $6k$.}
    \label{fig:TC_qubiteffic_6k}
\end{figure*}
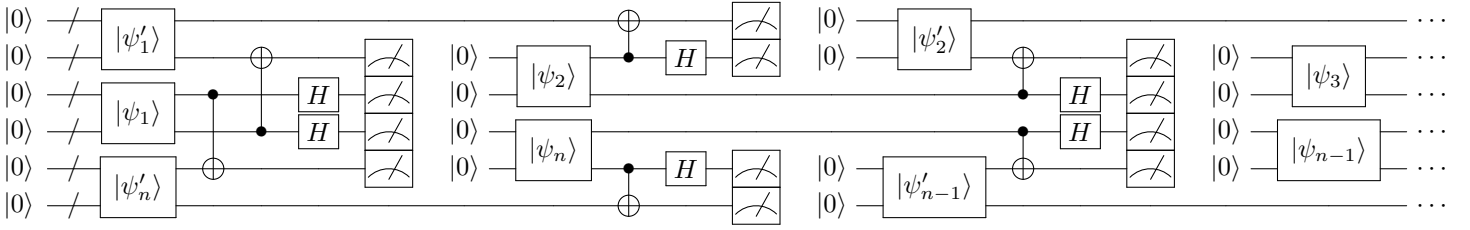
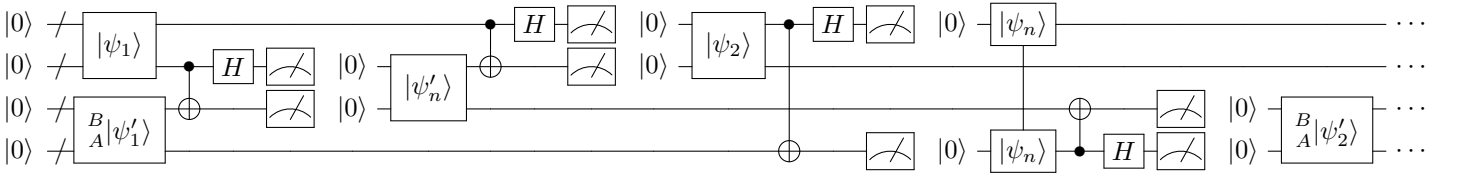
\begin{figure*}[t]
	\begin{equation*}
	\Qcircuit @C=0.5em @R=0.0em @!R {
        \lstick{\ket{0}} & \slash\qw & \multigate{1}{\ket{\psi_{1}}}						& \qw 		& \qw 		& \qw 		& \qw 												& \qw 						& \ctrl{1} 	& \gate{H} 	& \meter 	& \push{\rule{.2em}{0em}\ket{0}\rule{.2em}{0em}} 	& \multigate{1}{\ket{\psi_{2}}} & \ctrl{3} 	& \gate{H} 	& \meter & \push{\rule{.2em}{0em}\ket{0}\rule{.2em}{0em}} 	& \gate{\ket{\psi_{n}}}		& \qw		& \qw		& \qw		& \qw											 & \qw								 			& \push{\rule{.2em}{0em}\cdots\rule{.2em}{0em}}\qw	 \\
        \lstick{\ket{0}} & \slash\qw & \ghost{\ket{\psi_{1}}} 								& \ctrl{1} 	& \gate{H} 	& \meter 	& \push{\rule{.2em}{0em}\ket{0}\rule{.2em}{0em}} 	& \multigate{1}{\ket{\psi'_{n}}} 	& \targ 	& \qw 		& \meter 	& \push{\rule{.2em}{0em}\ket{0}\rule{.2em}{0em}} 	& \ghost{\ket{\psi_{2}}} 		& \qw 		& \qw		& \qw	 & \qw 												& \qw\qwx			& \qw		& \qw		& \qw		& \qw											 & \qw											& \push{\rule{.2em}{0em}\cdots\rule{.2em}{0em}}\qw	 \\
        \lstick{\ket{0}} & \slash\qw & \multigate{1}{\prescript{B}{A}{\ket{\psi'_{1}}}} 	& \targ 	& \qw 		& \meter 	& \push{\rule{.2em}{0em}\ket{0}\rule{.2em}{0em}} 	& \ghost{\ket{\psi'_{n}}} 		& \qw 		& \qw 		& \qw 		& \qw 												& \qw 					& \qw 		& \qw		& \qw	 & \qw												& \qw\qwx			& \targ		& \qw		& \meter 	& \push{\rule{.2em}{0em}\ket{0}\rule{.2em}{0em}} & \multigate{1}{\prescript{B}{A}{\ket{\psi'_{2}}}}	& \push{\rule{.2em}{0em}\cdots\rule{.2em}{0em}}\qw	\\
        \lstick{\ket{0}} & \slash\qw & \ghost{\prescript{B}{A}{\ket{\psi'_{1}}}} 			& \qw 		& \qw 		& \qw 		& \qw  												& \qw 						& \qw 		& \qw 		& \qw 		& \qw 												& \qw 					& \targ 	& \qw	 	& \meter & \push{\rule{.2em}{0em}\ket{0}\rule{.2em}{0em}} 	& \gate{\ket{\psi_{n}}}\qwx	& \ctrl{-1} & \gate{H} 	& \meter 	& \push{\rule{.2em}{0em}\ket{0}\rule{.2em}{0em}} & \ghost{\prescript{B}{A}{\ket{\psi'_{2}}}}			& \push{\rule{.2em}{0em}\cdots\rule{.2em}{0em}}\qw	
	}
	\end{equation*}
	\caption{The \qeTCTfourklong{}. Some of the copies of $\ket{\psi}$ are prepared ``upside-down'', preparing the $B$ subsystem on the upper wire, and some copies are prepared using non-adjacent wires. The circuit depth is $\Theta(n\times(\tsp+1))$, about twice that of \qeTCTsixk{}, the effective depth is $3(\tsp+O(1)) = \Theta(\tsp+1)$, and the width is $4k$.}
	\label{fig:TC_qubiteff_4k}
\end{figure*}

Our second algorithm comes from the observation that in the \qeHTfourk{}, the third register stays idle after the first state preparation.
So, instead of preparing two copies simultaneously, we modify the algorithm to prepare one copy, reset the qubits associated with subsystem $B$, and reuse them to prepare successive copies. This saves $k$ qubits.
This algorithm is given in Fig.~\ref{fig:H_qubiteff_3k}.
Here, the circuit width is $3k+1$ qubits. We refer to this as the \qeHTthreek{}.

Our two qubit-efficient versions differ only slightly. The second version requires $k$ fewer qubits than the first one. This savings come at the cost that the second wire will have to wait longer before gates are applied, exposing it to more thermal noise. The length of the extra wait depends on how long state preparation takes, but compared to the depth of the $n-1$ other state preparations, the effect should be negligible. After the first two state preparations, the circuits are effectively the same.

Next, we compare the two qubit-efficient versions to the original \HT{} algorithm (Fig.~\ref{fig:JSTalg_HTestwithPermutation}). 
First, all of the circuits have the same number of gates and measurements, so we expect gate and readout errors to affect them similarly. 
If the fidelity of qubit reinitialization, i.e. qubit reset,
is significantly worse than the fidelity of initialization in the beginning of computation, the qubit-efficient algorithms will have a disadvantage.
The depth of the original algorithm is $\tsp{}+\Theta(kn)$ while the depths of the two qubit-efficient algorithms are $\Theta\left(n\times (\tsp{}+k) \right)$. 
Thus, when $\tsp{}$ is small, the original and new algorithms have similar depth. 
Fortunately, even short-depth circuits have the potential to prepare many interesting states; indeed, the recent quantum supremacy experiment by~\cite{arute2019quantum} used 53-qubit circuits with just forty layers of gates.
These observations suggest that our new algorithms may perform similarly to the original algorithm, given small \tsp{}, even as they achieve asymptotically lower circuit width.

\subsection{Qubit-Efficient \TCT}
\label{sec:QET}

In the \TCT{} (Fig.~\ref{fig:tc_alg}), each copy of the state interacts with two other copies of the state, one via its $A$ subsystem and one via its $B$ subsystem. After these interactions and, in the case of the $A$ subsystem, a Hadamard gate, the registers containing that copy can be measured, reset, and reused.
Therefore, we just need enough qubits to maintain three copies of the state.
However, we must be careful, since while the \HT{} did not require any particular ordering of the $n$ copies of $\ket{\psi}$, the \TCT{} does.
Fortunately, the \TCT{} is structured such that simply following a greedy strategy of preparing whichever copy is needed to interact with the current longest-lived copy is sufficient.

Our first qubit-efficient variant is given in Fig.~\ref{fig:TC_qubiteffic_6k}. The circuit width is $6k$ qubits, so we refer to this algorithm as the \qeTCTsixk{}.
Recall that we refer to the first $n$ copies of the state by $\ket{\psi_i}$ and to the second $n$ copies (which are acted on by the permutation operator) by $\ket{\psi_i'}$.

To further reduce the number of qubits, we observe that it is unnecessary to simultaneously prepare both copies needed by the current one. For example, after preparing $\ket{\psi_1}$, it is sufficient to first prepare $\ket{\psi_1'}$, interact the $B$ subsystems of those copies, and \emph{then} prepare $\ket{\psi_n'}$ and interact the $A$ subsystems. 
The register containing the $B$ subsystem of $\ket{\psi_1'}$ can be measured, reset, and reused to prepare $\ket{\psi_n'}$. 
In this way, four such registers is sufficient.
Our second variant is given in Fig.~\ref{fig:TC_qubiteff_4k}.
The circuit width is $4k$, and we refer to it as the \qeTCTfourk{}.

Our two qubit-efficient variants are similar.
The second version uses $2k$ fewer qubits. In both versions, half of the wires are measured quickly, after just $\tsp{}+O(1)$ timesteps. 
While the remaining wires in the first algorithm are used for $2\tsp{}+O(1)$ timesteps between initialization and measurement, the wires in the second algorithm must be maintained for $3\tsp{}+O(1)$ time.
So, the second algorithm may suffer from thermal noise more than the first. %

Next, we compare the two qubit-efficient versions to the original \TCT{} algorithm. First, all the circuits have the same number of gates and measurements, so we expect gate and readout errors to affect them similarly. 
If the fidelity of qubit reinitialization, i.e. qubit reset, is significantly worse than the fidelity of initialization in the beginning of computation, the qubit-efficient algorithms will have a disadvantage.
The original \TCT{} has  depth $\tsp{}+O(1)$, while our qubit-efficient variants have depth $\Theta\left(n\times (\tsp{}+1)\right)$ (note the constant term is only asymptotic, like $O(1)$, rather than a literal 1).
Based on this observation, the qubit-efficient versions might appear like they should perform significantly worse in the presence of noise. 

Given the results of~\cite{subasi2018entanglement} demonstrating that the \TCT{} is more noise-resilient than the \HT{}, we expect that each of the variants of the \TCT{} should outperform their \HT{} analogs, e.g. we expect the \qeTCTsixk{} to outperform the \qeHTfourk{}. However, it is unclear a priori whether the qubit-efficient variants of the \TCT{} will still outperform the original \HT{}.

\section{Numerical simulations}
\label{sec:numerics}

In this section, we test the performance of our qubit-efficient algorithms for entanglement spectroscopy and compare them to the original versions.
The most significant observation from our results is that our qubit-efficient algorithms perform similarly to the originals in the presence of noise.

We use IBM's Qiskit~\cite{Qiskit} and QASM simulator to numerically simulate noisy quantum circuits. 
Our simulations include thermal relaxation and decoherence error, readout error, and gate noise in the form of depolarizing and Pauli errors. See the Appendix for details and parameters.
These simulations and selection of noise parameters are independent of the experimental results in the next section.
The postselection methods introduced by \cite{linke2018measuring,subasi2018entanglement} for improving the accuracy of the \HT{} and \TCT{} apply in the same way to our qubit-efficient variants, but because we expect them to affect the old and new algorithms similarly, we do not implement postselection here.

The number of qubits we can simulate is limited by memory, and the circuit widths of the original algorithms scale with $n$ and $k$ while the widths of our new algorithms scale only with $k$. 
So, in order to simulate the prior algorithms for many values of $n$, we restrict our simulations to $k=1$, which corresponds to two-qubit states $\ket{\psi}$ and single-qubit density matrices $\rho_A$. 
In this case, knowing $\tr(\rho_A^n)$ for $n=2$ is sufficient to reconstruct the entire entanglement spectrum. Although the values for $n>2$ are redundant, we compute them in order to assess the performance of the algorithms.

\begin{figure}[t]
    \begin{equation*}
    \Qcircuit @C=0.5em @R=0.0em @!R {
        \lstick{A}  & \qw &	\gate{H} 		 									& \ctrl{1} & \qw	\\
        \lstick{B}	& \qw &	\gate{U_2(\theta\!-\!\frac{\pi}{2},\frac{\pi}{2})} 	& \targ    & \qw
	}
    \end{equation*}
    \caption{State preparation circuit. The $U_2(\phi,\lambda)$ gate has the matrix form $\frac{1}{\sqrt{2}}((1, -e^{i \lambda}),(e^{i\phi},e^{i(\phi+\lambda)}))$.}
    \label{fig:stateprep}
\end{figure}
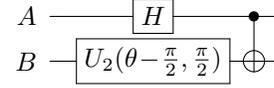

For each $n$, we generate twenty quantum states with varying levels of entanglement ranging from product states to maximally entangled using the circuit in Fig.~\ref{fig:stateprep}. 
We choose the twenty angles $\theta$ therein such that the associated $\tr(\rho_{A}^n)$ are evenly spaced from the minimum to the maximum possible values, from $2^{1-n}$ (fully mixed) to $1$ (pure state).

\begin{figure*}[t]
    \centering
    \includegraphics[width=\textwidth]{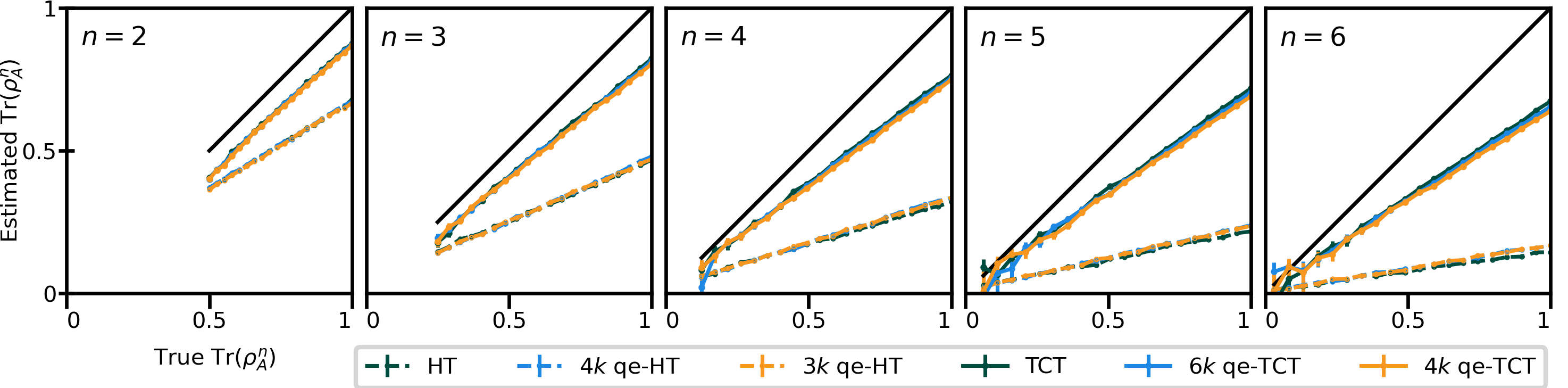}
    \caption{(Color) Numerical simulations of prior algorithms and our variants estimating twenty values of $\tr(\rho_A^n)$ under simulated hardware noise for several values of $n$. Noise parameters are in the Appendix. True values are on the horizontal axis, estimated values on the vertical. 
    The topmost line shows ideal results. 
    All points are marked.
    Error bars, some of which are smaller than the lines, are based on expected statistical noise only.
    The upper group of lines are the variants of the \TCT{}. The lower group of dashed lines are the variants of the \HT{}. 
    We compare the slopes of these lines in Fig.~\ref{fig:slopePlot_all_n2to6}.}
    \label{fig:linearPlots_allAlgs}
\end{figure*}

After simulating the algorithms for all twenty states corresponding to a particular $n$, we first plot the values of the ideal $\tr(\rho_A^n)$ versus the value estimated by each quantum algorithm. Fig.~\ref{fig:linearPlots_allAlgs} shows these plots for simulations of all algorithms, including original algorithms and our qubit-efficient algorithms, for $n=2$ to $n=6$.
Note that an ideal set of results would lie on a straight line from $(2^{1-n},2^{1-n})$ to $(1,1)$ with slope equal to one. 
Our results deviate from this line due both to simulated hardware noise and to statistical noise due to finite sampling.
Intuitively, random hardware noise leads pure states to appear more mixed, leading the results to concentrate about a flatter line, and statistical noise causes the results to deviate about that line.
Observe that as long as the data concentrate about some line, it is easier to confidently identify a state as more or less mixed based on the algorithm's estimate for $\tr(\rho_A^n)$ when the slope of the line is closer to one, i.e. when the line is steeper.
This is in contrast to an error in the vertical intercept of a line, which can be corrected by learning the error and shifting future results.
Therefore, we characterize the performance of the algorithms by their slopes.
For each value of $n$ that we tested, we compute the slope of each line in plots like in Fig.~\ref{fig:linearPlots_allAlgs} using a linear regression. We plot the values of $n$ versus those computed slopes.

\begin{figure}[h]
    \centering
    \includegraphics[width=\columnwidth]{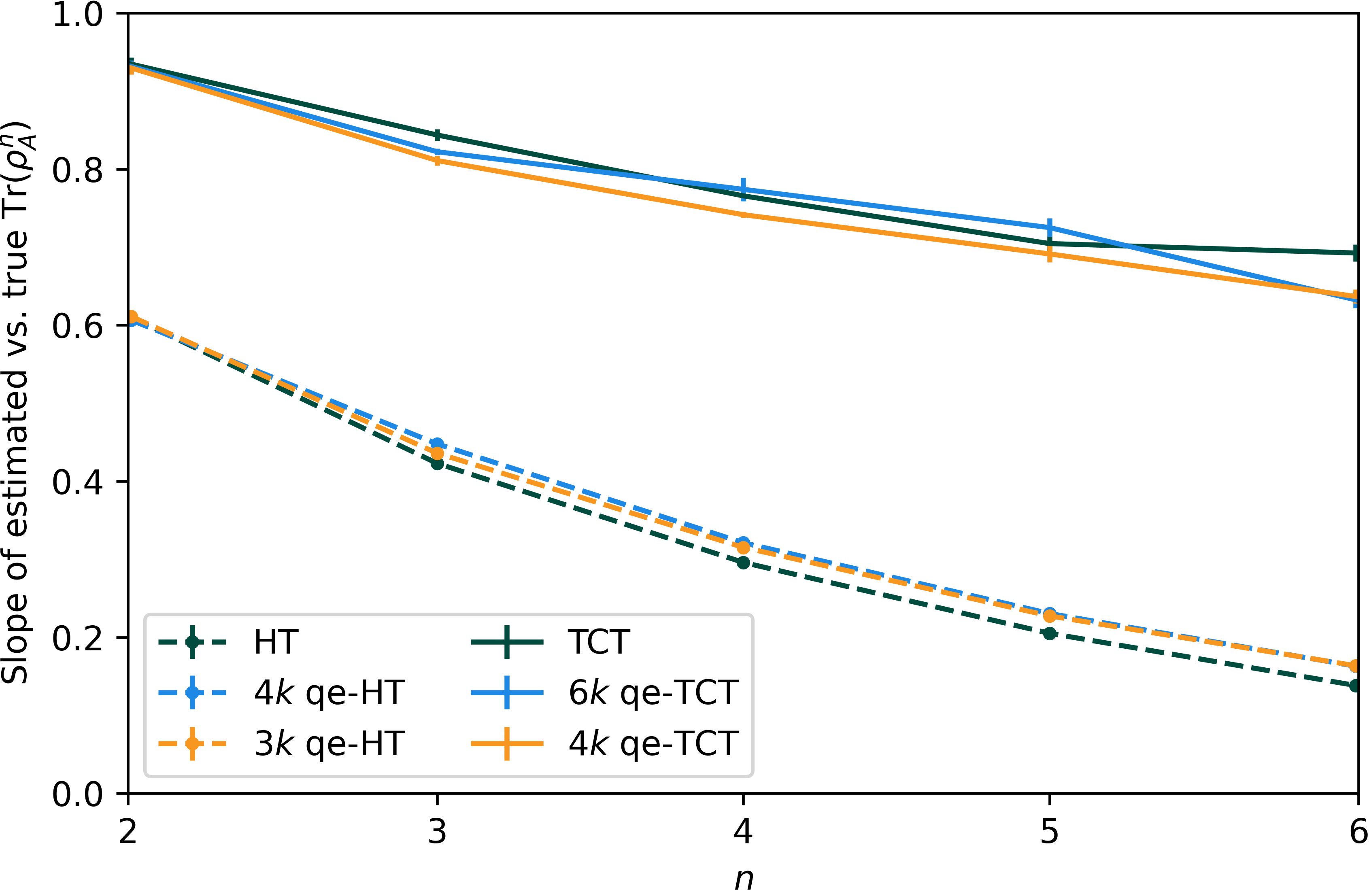}
    \caption{(Color) Comparison of the performance of all \HT{} and \TCT{} variants in the presence of simulated noise for $n=2$ to $n=6$. 
    Values of $n$ are on the horizontal axis, slopes computed by a linear regression on the corresponding lines in Fig.~\ref{fig:linearPlots_allAlgs} are on the vertical axis.
    Ideal results would produce a horizontal line at height $1$.
    Error bars, some of which are smaller than the lines, are based on the quality of the linear fit for the corresponding data in Fig.~\ref{fig:linearPlots_allAlgs}.
    The upper group of lines are the \TCT{} variants. The lower group of dashed lines, with marked points, are the \HT{} variants.}
    \label{fig:slopePlot_all_n2to6}
\end{figure}

Results for all algorithms, including original algorithms and our qubit-efficient versions, for $n=2$ to $n=6$  are given in Fig.~\ref{fig:slopePlot_all_n2to6}.
Note that a noiseless implementation would have slope equal to one for all $n$.
Decreasing values indicate an algorithm's performance degrading for larger values of $n$.
We were limited to $n=6$ by the \TwoCopyTest{}, for which the circuit width scales as $4kn$; simulating $28$ qubits was impractical due to time-constraints, and $32$ qubits would be impractical due to memory constraints.
In contrast, the number of qubits required for our qubit-efficient algorithms is independent of $n$, so we are able to simulate these algorithms for much larger values of $n$. 

Results for the qubit-efficient algorithms for $n=2$ to $n=20$ are given in Fig.~\ref{fig:slopePlot_qe_n2to20}. The noise strength is reduced compared to the previous simulations (see the Appendix for details). 

The most significant observation from these results is that our qubit-efficient algorithms perform similarly to the original variants.
In Fig.~\ref{fig:slopePlot_all_n2to6}, the qubit-efficient variants of the \HT{} perform very similarly to the original algorithm.
In fact, the original \HT{} performs slightly worse than the qubit-efficient variants, likely due to Qiskit's ``as soon as possible'' gate scheduling which prepares copies of $\ket{\psi}$ for the original \HT{} earlier than is optimal.
As we stated previously, we expected the qubit-efficient variants to perform similarly to the original when \tsp{} is small. 
In the case of the \TCT{}, the qubit-efficient variants suffer almost no degradation compared to the original algorithm.
For both the \HT{} and \TCT{}, the wider, lower-depth qubit-efficient variants perform better than the corresponding lower-width algorithms.
We note that, as explored further in~\cite{subasi2018entanglement}, the \TCT{} and its variants are more susceptible to statistical noise than the \HT{}. The \TCT{} is most affected by statistical noise when estimating small values of $\tr(\rho_A^n)$, which is the case for highly entangled states $\ket{\psi}$ and exasperated by large powers $n$; this is visible in Fig.~\ref{fig:linearPlots_allAlgs}.

In Fig.~\ref{fig:slopePlot_qe_n2to20}, simulating larger $n$, the two qubit-efficient variants of \HT{} continue to perform almost identically, as expected.
For the \TCT{} variants, the \qeTCTsixk{} slightly outperforms the \qeTCTfourk{}.
Notably, both the qubit-efficient variants of the \TCT{} still appear to produce meaningful results when $n=20$ (as good as the \qeHT{} when $n=8$).

\begin{figure}[h]
    \centering
    \includegraphics[width=\columnwidth]{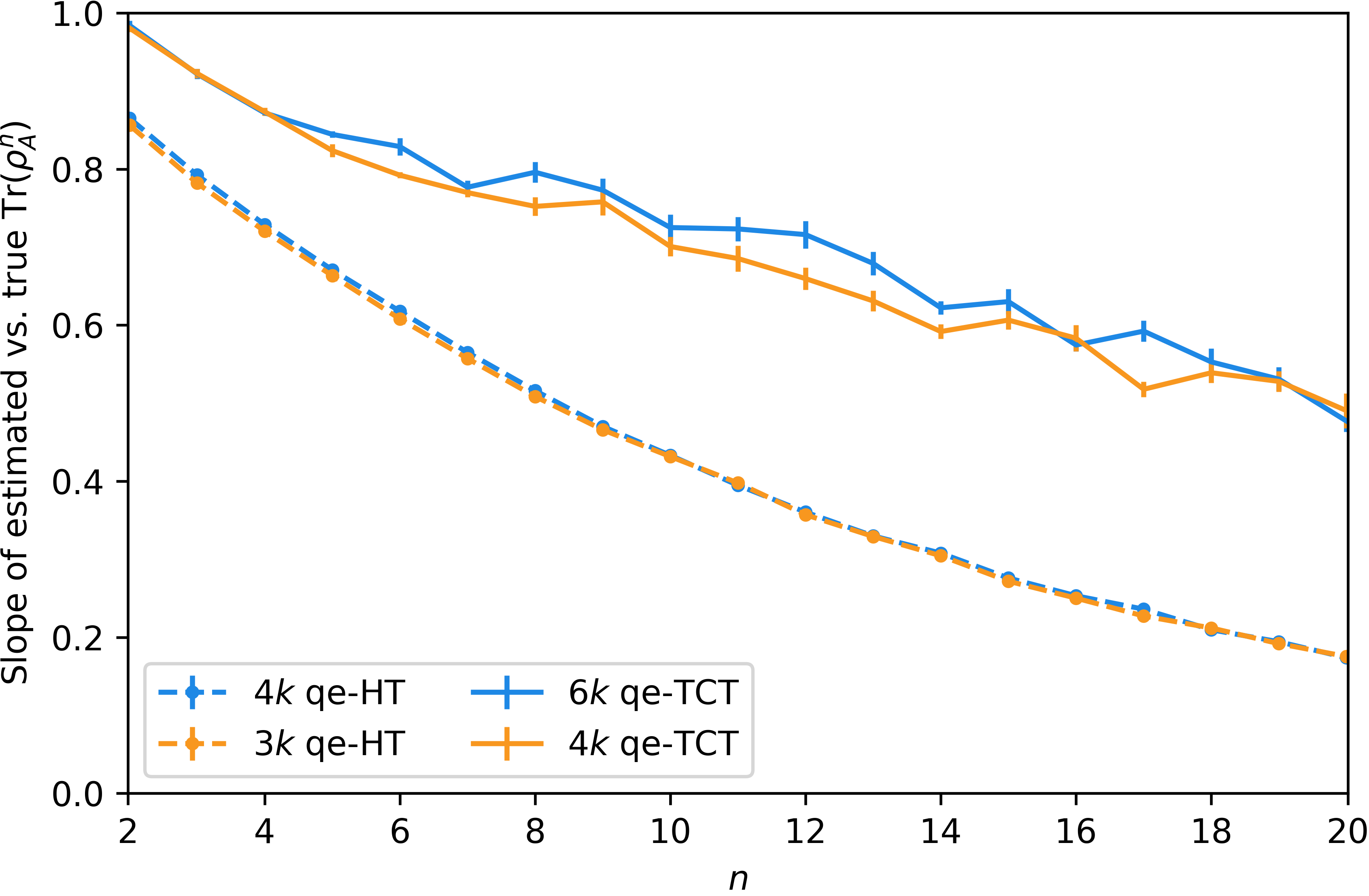}
    \caption{(Color) Comparison of the performance of our qubit-efficient variants of the \HT{} and \TCT{} for $n=2$ to $n=20$. The noise strength is reduced here compared to the previous simulations (see the Appendix for details).}
    \label{fig:slopePlot_qe_n2to20}
\end{figure}

\section{Experimental demonstration on Honeywell System Model H0}
\label{sec:experiment}

In this section, we report the results of testing one of our qubit-efficient algorithms on the Honeywell System Model H0~\cite{pino2020demonstration}. 
We were able to estimate $\tr(\rho_A^n)$ for larger $n$ than would have been possible on the device using any previous algorithm, and our results correctly distinguish more and less entangled states.

This quantum computer is a trapped-ion quantum charge-coupled device architecture; for details, see~\cite{pino2020demonstration}. 
At the time of access (September 2020), the device supported six qubits and supported mid-circuit measurements and qubit resets. 
In order to test the widest variety of parameters possible given our limited time on the device, we chose to test one qubit-efficient algorithm, choosing the one which performed best in simulations: the \qeTCTsixk{} (Fig.~\ref{fig:TC_qubiteffic_6k}).

As in our numerical simulations, we set $k=1$, corresponding to two-qubit states $\ket{\psi}$ and one-qubit $\rho_A$.
We prepare three states with varying levels of bipartite entanglement using the state preparation circuit of Fig.~\ref{fig:stateprep}, setting the angle $\theta$ therein to $\theta=1.33,1.05,0.87$. 
Because the \TCT{} is more sensitive to statistical noise when estimating smaller values of $\tr(\rho_A^n)$, corresponding to more mixed $\rho_A$, and because we had only a limited number of runs available, we chose these states to be closer to pure than to fully mixed. 

For each of the three states, we run the \qeTCTsixk{} for $n=2,\dots,7$ for 1,000 runs.
Note that given six qubits, the original \TCT{} would not fit on the device even for $n=2$.
Each circuit was sent via the HQS API, specified using operations $U_2, \CNOT{}$, Measure, and Reset ($U_2$ defined in Fig.~\ref{fig:stateprep}). 
From there, each circuit was compiled to the device's native gate set, including standard optimization according to Honeywell's software stack, and submitted to the device.
Circuits were sent in batches, with calibration performed within and between each batch.

\begin{figure}[h]
    \centering
    \includegraphics[width=\columnwidth]{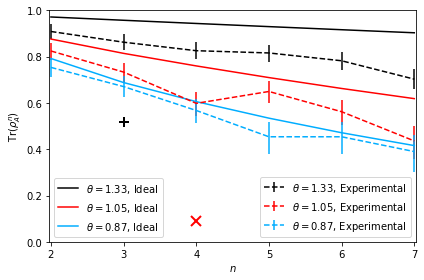}
    \caption{(Color) Results from our experimental test of the \qeTCTsixk{} on the Honeywell System Model H0. Values of $n$ are on the horizontal axis, $\tr(\rho_A^n)$ on the vertical. Three different states $\rho_A$ corresponding to different values of $\theta$ were used. For each state, there is a solid line of true values and a dashed line of experimental results. 
    Tests were repeated for two data points because the results were outliers; the initial points, one for $\theta=1.33$ and one $\theta=1.05$, are respectively marked by `$+$' and `$\times$'.
    Error bars are based on expected statistical noise.}
    \label{fig:honeywellPlot}
\end{figure}

Results are shown in Fig.~\ref{fig:honeywellPlot}.
Rather than comparing several algorithms, here we test the performance of the \qeTCTsixk{} on several different inputs. 
For each of the three states, we plot the values of $n$ versus the estimates for $\tr(\rho^n)$.

After receiving the results from our tests, we found that two data points, for $\theta=1.33,n=3$ and for $\theta=1.05,n=4$, were outliers compared to the rest of the data. Honeywell offered to rerun these tests. Both the initial and second points are shown in Fig.~\ref{fig:honeywellPlot}.

Because of noise, the results from our tests are insufficient to recover the true, analytical values of $\tr(\rho_A^n)$. 
However, results for each of the three states are clearly distinguishable from each other and are correctly ordered according to their degree of entanglement. 
The data is remarkably smooth across varying $n$, with simulations predicting more varied outcomes and with these tests using only 1,000 runs versus the simulations in Fig.~\ref{fig:linearPlots_allAlgs} using 100,000.
Although we only tested the algorithm on states closer to pure than fully mixed (recall the minimum value of $\tr(\rho_A^n)$ is $2^{1-n}$), the results appear promising for more entangled states. They also suggest that tests with larger values of $n$ should produce results along a similar trend.

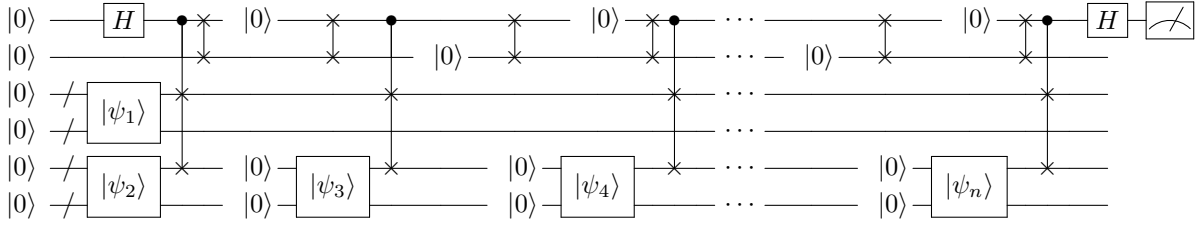
\begin{figure*}[t]
    \begin{equation*}
    \Qcircuit @C=0.7em @R=0.0em @!R {
	\lstick{\ket{0}}    & \qw       & \gate{H}  					& \ctrl{2}  	& \qswap    	& \push{\rule{.6em}{0em}\ket{0}\rule{0em}{0em}}\qw		& \qswap						& \ctrl{2}   	& \qw												& \qswap											& \push{\rule{.6em}{0em}\ket{0}\rule{0em}{0em}}\qw 		& \qswap		& \ctrl{2} 		& \qw		& \push{\rule{.2em}{0em}\cdots\rule{.2em}{0em}}\qw 	& \qw												& \qswap											& \push{\rule{.6em}{0em}\ket{0}\rule{0em}{0em}}\qw 		& \qswap		& \ctrl{2} 		& \qw	& \gate{H} & \meter \\
	\lstick{\ket{0}}	& \qw       & \qw							& \qw\qwx		& \qswap\qwx	& \qw 													& \qswap\qwx					& \qw\qwx		& \push{\rule{.6em}{0em}\ket{0}\rule{0em}{0em}}\qw	& \qswap\qwx    									& \qw													& \qswap\qwx 	& \qw\qwx		& \qw		& \push{\rule{.2em}{0em}\cdots\rule{.2em}{0em}}\qw	& \push{\rule{.6em}{0em}\ket{0}\rule{0em}{0em}}\qw	& \qswap\qwx    									& \qw													& \qswap\qwx 	& \qw\qwx		& \qw	& \qw \\
	\lstick{\ket{0}}    & \slash\qw & \multigate{1}{\ket{\psi_{1}}}	& \qswap    	& \qw 			& \qw 													& \qw 							& \qswap     	& \qw												& \qw 												& \qw 													& \qw 			& \qswap 		& \qw		& \push{\rule{.2em}{0em}\cdots\rule{.2em}{0em}}\qw 	& \qw												& \qw 												& \qw 													& \qw 			& \qswap 		& \qw  	& \qw \\
	\lstick{\ket{0}}    & \slash\qw & \ghost{\ket{\psi_{1}}}       	& \qw \qwx  	& \qw 			& \qw 													& \qw       					& \qw\qwx   	& \qw												& \qw 												& \qw 													& \qw 			& \qw \qwx 		& \qw		& \push{\rule{.2em}{0em}\cdots\rule{.2em}{0em}}\qw	& \qw												& \qw 												& \qw 													& \qw 			& \qw \qwx 		& \qw	& \qw \\
	\lstick{\ket{0}}    & \slash\qw & \multigate{1}{\ket{\psi_{2}}}	& \qswap\qwx 	& \qw 			& \push{\rule{.6em}{0em}\ket{0}\rule{0em}{0em}}\qw     	& \multigate{1}{\ket{\psi_{3}}} & \qswap\qwx	& \qw												& \push{\rule{.6em}{0em}\ket{0}\rule{0em}{0em}}\qw 	& \multigate{1}{\ket{\psi_{4}}} 						& \qw 			& \qswap \qwx 	& \qw 		& \push{\rule{.2em}{0em}\cdots\rule{.2em}{0em}}\qw	& \qw												& \push{\rule{.6em}{0em}\ket{0}\rule{0em}{0em}}\qw 	& \multigate{1}{\ket{\psi_{n}}} 						& \qw 			& \qswap \qwx 	& \qw	& \qw \\
	\lstick{\ket{0}}    & \slash\qw & \ghost{\ket{\psi_{2}}}       	& \qw       	& \qw 			& \push{\rule{.6em}{0em}\ket{0}\rule{0em}{0em}}\qw 		& \ghost{\ket{\psi_{3}}}        & \qw		   	& \qw												& \push{\rule{.6em}{0em}\ket{0}\rule{0em}{0em}}\qw 	& \ghost{\ket{\psi_{4}}} 								& \qw 			& \qw			& \qw 		& \push{\rule{.2em}{0em}\cdots\rule{.2em}{0em}}\qw	& \qw												& \push{\rule{.6em}{0em}\ket{0}\rule{0em}{0em}}\qw 	& \ghost{\ket{\psi_{n}}} 								& \qw 			& \qw			& \qw	& \qw
}
    \end{equation*}
    \caption{An alternative implementation of the \qeHT{} that uses an extra qubit to disguise how long the ancilla qubit is required to remain coherent.}
    \label{fig:H_qubiteff_4k_highdepth_nonObvious}
\end{figure*}
\section{Effective circuit depth}
\label{sec:effecDepth}

In this section, we introduce a generalization of circuit depth which is more useful for circuits using qubit resets, which we call effective depth.
The depth of a circuit is defined as the number of timesteps assuming that gates can be applied in parallel, or equivalently as the maximum length of a path from the input to the output. 
Circuit depth is often used to quantitatively judge how susceptible a quantum computation will be to thermal decoherence and relaxation noise.
Intuitively, the higher a circuit's depth, the more time during which the circuit may be affected by noise.
This is especially relevant in the NISQ era, as coherence times remain a primary limiting factor on tractable problem sizes.
However, depth is only a heuristic for judging noise-resilience. Circuits may be affected by various sources of noise besides thermal noise, and comparing the depths of two circuits does not perfectly predict the relative performance of the circuits even when the noise model is restricted to thermal noise.
For example, circuits which produce highly entangled states will be significantly more affected by decoherence than circuits which remain in computational basis states (entirely classical information) even when those circuits have the same depth.
Nevertheless, considered alongside other factors, circuit depth is a convenient, often-used tool for assessing quantum algorithms.

In the setting of circuits that use qubit resets, circuit depth is no longer useful for assessing noise resilience.
Consider, for example,
that the depth of the original \TCT{} is $\tsp{}+O(1)$ while the depths of the qubit-efficient versions are $\Theta(n\times(\tsp{}+1))$, an asymptotic increase.
But, as shown numerically in Section~\ref{sec:numerics}, the algorithms perform similarly in the presence of noise.
Circuit depth judges circuits with resets too harshly.
Anticipating increased use of qubit resets, we would like a measure which incorporates their presence.

Defining such an attribute is subtle. A naive idea for a depth-like predictor of noise-resilience for circuits with qubit resets might be the largest amount of time between resets of any particular qubit. %
However, consider the alternative implementation of the \qeHT{} shown in Fig.~\ref{fig:H_qubiteff_4k_highdepth_nonObvious}. This implementation utilizes two ancilla qubits where our previous implementations used one, making frequent swaps between the two ancilla qubits. It is designed to obfuscate the long time for which the ancilla qubit in the \qeHT{} must be kept coherent. This naive measure would rate this circuit as $\Theta(k)$, and with further changes this could be made $O(1)$. Clearly though, information stored in the ancilla is just as exposed to thermal noise in this new circuit as in the other qubit-efficient circuits. %

Instead, our definition is inspired by the idea of information flow and locality.
At a high level, quantum information 
is only transferred between qubits when multi-qubit gates are applied.\footnote{We consider a level of abstraction that ignores potential crosstalk.}
In particular, the corruption of quantum information due to noise on one register cannot propagate to another register except through future multi-qubit interactions. 
These ideas are considered further in~\cite{schumacher2005locality,schumacher2012isolation}. 
By focusing on information flow, the shortcomings of traditional circuit depth for circuits using resets can be eliminated.

We define the \emph{effective depth} of a circuit to be the maximum length of a path along which there is information flow. Equivalently, it is the maximum number of timesteps for which some quantum information is propagated. 
Such directed paths can be constructed by beginning from any qubit (re)initialization, following the qubit, optionally crossing from one qubit to another when there is a two-qubit gate between them, and terminating when there is a reset or when the last operation is reached;\footnote{A similar but distinct construction is used in~\cite{broadbent2009parallelizing} with the different motivation of parallelizing circuits.} the longest path which can be formed in this way gives the effective depth.

To justify our definition, consider the following. First, effective depth reduces to the standard definition of depth for circuits which do not use resets.
Second, observe that the graph of paths stemming from any set of qubit initializations can effectively be viewed as a subcircuit. 
Then, an equivalent definition of effective depth is the maximum (standard) depth of such a subcircuit.
From this perspective,  effective depth is a natural extension of depth, rating a circuit with resets according to the depth of its largest complete subcircuit.
Third, although effective depth is not a perfect tool, it is a heuristic which provides a worst-case assessment just as standard circuit depth does. %
The length of the longest path may be unusually long compared to the rest of the paths, %
some multi-qubit gates may transfer information asymmetrically, or the distribution of inputs may mean certain paths are more significant than others: 
these factors should be considered alongside effective depth just as additional factors are needed alongside traditional depth.
Finally, effective depth sensibly rates all of the circuits in this article, as we discuss next.

Since effective depth reduces to standard depth for circuits which do not use resets, %
the effective depth of \HT{} is $\tsp{}+\Theta(nk)$ and the effective depth of \TCT{} is $\tsp{}+O(1)$. 
For the qubit-efficient variants of \HT{}, the ancilla qubit leads to effective depths the same as their depths, $\Theta(n\times(\tsp+k))$, which asymptotically matches that of \HT{} when $\tsp{}$ is small. 
Now, for the qubit-efficient \TCT{} variants, standard circuit depth is insufficient to explain our numerical results. %
The effective depth of \qeTCTsixk{} is $2(\tsp+O(1))$ and the effective depth of \qeTCTfourk{} is $3(\tsp+O(1))$. 
These values asymptotically match the effective depth of the original \TCT{}, $\tsp+O(1)$, 
which helps explain why our \qeTCT{}s perform just as well as the original \TCT{}: their effective depths are the same.\footnote{Intuitively, the \qeTCTfourk{} should experience three times more thermal noise (two times for the \qeTCTsixk{} algorithm) than the original \TCT{} because its effective depth is three (two) times greater. We have tested this intuition using numerical simulations with only thermal noise, multiplying the gate times for the original \TCT{} by three, and found it correct.}
Finally, effective depth assigns the same value to the contrived \qeHT{} circuit (Fig.~\ref{fig:H_qubiteff_4k_highdepth_nonObvious}) as standard depth, thus avoiding the potential pitfall of assessing this circuit too gently.
Our numerical results in  Section~\ref{sec:numerics} are consistent with all these observations.

\section{Discussion}
\label{sec:discuss}

In this work, we introduced new qubit-efficient algorithms for performing entanglement spectroscopy via computing $\tr(\rho_A^n)$ that use qubit resets to achieve asymptotically lower width than previous algorithms.
Our numerical results show that the performance of our algorithms is only slightly degraded by noise even as they save a significant number of qubits.
First, the \qeHTlong{} requires as few as $3k+1$ qubits and achieves similar performance to the original \HT{} algorithm; we expect this to hold given small state preparation time $\tsp$. Second, and in particular, the \qeTCTlong{} requires as few as $4k$ qubits while achieving similar performance to the original \TCT{} algorithm, and we expect this to hold in general.
Our algorithms demonstrate the usefulness of the as yet understudied tool of qubit resets.

Just as the \HT{} algorithm of~\cite{johri2017entanglement} may be better than the \TCT{} for the case of $n=2$ (i.e. the \SwapTest{}), the original \TCT{} algorithm of~\cite{subasi2018entanglement} may remain preferable to our new variants for small powers $n$.
Our approach is preferable for values of $k$ and $n$ where at least $4kn$ qubits are unavailable or when a smaller circuit width is desired. 

To demonstrate the practicality of our qubit-efficient algorithms, we experimentally implemented the \qeTCTsixk{} for $k=1$ and $n=2,\dots,7$ on the Honeywell System Model H0, which supported (at the time of implementation) six qubits. 
As a comparison, the original \TCT{} would require 8 qubits for $n=2$ and so could not be run for any $n$, while the original \HT{} would require 5 qubits for $n=2$ and could not be run for any larger $n$.
Although the results of the experiment are too noisy to immediately recover the spectrum, they successfully differentiate and rank states with different amounts of entanglement, which could be useful for quantum simulation applications in the near future.

Traditional circuit depth is insufficient for assessing our algorithms or future algorithms using qubit resets.
In contrast, effective depth justifies the performance of our new algorithms; for example, the qubit-efficient variants of the \TCT{} have the same asymptotic effective depth as the original \TCT{}.
Our definition is a simple and useful heuristic for predicting noise-resilience, just as traditional circuit depth is for circuits without resets.
Notably, when there are no qubit resets present, effective depth reduces to standard circuit depth.
Effective depth will be a useful tool in the future design and analysis of qubit-efficient algorithms and it should be preferred over circuit depth for describing circuits with qubit resets.

    Quantum error mitigation techniques developed for use in NISQ devices~\cite{endo2021hybrid} may improve the performance of our algorithms.
    Notably, postselection strategies for the \HT{}~\cite{linke2018measuring} and the \TCT{}~\cite{subasi2018entanglement} can also be used with the corresponding qubit-efficient algorithms (these methods fit into the broader framework of symmetry verification~\cite{bonet2018low}). 
    These postselection strategies generally shift estimates of $\tr\left(\rho_A^n\right)$ upward by a constant independent of the input $\rho_A$ %
    (see Fig.~6 of~\cite{subasi2018entanglement}). 
    This is indeed useful for obtaining more accurate estimates. However, 
    not having a particular application in mind, in this work we decided to compare the performance of various algorithms by their ability to distinguish varying degrees of entanglement (see Sec.~\ref{sec:numerics}). Postselection does not seem to improve this ability due to the uniform improvement in estimates.
    Extrapolation~\cite{li2017efficient,temme2017error} techniques may be particularly helpful in this regard.
    Throughout our tests, we observed that noise shifts the algorithms' estimates of $\tr\left(\rho_A^n\right)$ proportionally to the ideal value such that results plotted as in Fig.~\ref{fig:linearPlots_allAlgs} consistently remain linear; this effect can be leveraged by testing the algorithm on some known states and comparing the ideal and experimental outputs to extrapolate the effect of noise and correct for the error.
    We leave further improvement and error mitigation, which will depend on a range of factors including the particular hardware and inputs, for future work.
    
    Similarly, further work on analytically modeling the effect of various potential errors on our algorithms would help to improve their performance. The analysis of~\cite{foulds2021controlled} on the robustness of the \SwapTest{} would be a good starting point for analyzing the \HT{} and the \TCT{}.
    The \TCT{} is a special case of convolutional circuits recently studied by~\cite{anikeeva2021recycling}. 
    In addition to being convolutional, all of the variants of \TCT{} have constant depth and constant effective depth. These features of \TCT{} provide a framework for understanding its noise resilience.

Developing qubit-efficient algorithms will be critical in the NISQ era. 
Similar devices with fewer qubits tend to be less noisy than those with more qubits, so it is advantageous to be able to run an algorithm on the smallest quantum device possible. 
Given a particular device, carefully choreographing operations, qubits resets, and the resulting flow of information will help increase the size of the largest problems that can be solved.
Additionally, because these algorithms use fewer qubits, they will benefit from requiring fewer swaps to implement gates between arbitrary qubits on architectures with limited connectivity.
The performance could be further improved by designing special purpose devices optimized to run these algorithms. 
Ongoing work on compiling and optimizing quantum algorithms may enable automatically using qubit resets to reduce circuit width, as well as optimizing reset placement based on qubit connectivity and noise.

As shown in this work, entanglement spectroscopy is one application for which qubit-efficient algorithms are possible. 
Efficient characterization of the entanglement in quantum states will be useful in many areas. In particular, it is well-suited to the promising NISQ application of quantum simulation.
In this context, our qubit-efficient algorithms might be paired with quantum simulation methods which utilize qubit resets in order to reduce the necessary number of qubits, such as recent work on simulating correlated spin systems~\cite{fossfeig2020holographic}.
Our algorithms may also prove helpful in characterizing the performance of NISQ devices themselves.

Additional algorithms, known and future, may be implemented with fewer qubits using qubit resets.
Promising candidates include algorithms which are already low-depth and which have a structure such that registers generally do not require interaction with many other registers.

\section*{Acknowledgments}
We thank the Honeywell Quantum Solutions team and Andrew Potter for inspiring discussions.
We thank anonymous referees for helpful suggestions of additional details and improvements to the presentation.
Part of this work was completed while JY was a participant in the 2019 Quantum Computing Summer School at Los Alamos National Laboratory (LANL), sponsored by the LANL Information Science \& Technology Institute.
JY acknowledges support from a Vannevar Bush Faculty Fellowship from the Department of Defense, Opportunity No. N0001415RFO11.
YS acknowledges funding from LANL ASC Beyond Moore's Law project and the LDRD program at LANL. 
Los Alamos National Laboratory is managed by Triad National Security, LLC, for the National Nuclear Security Administration of the US Department of Energy under Contract No. 89233218CNA000001.

\appendix
\section{Numerical simulation details}

\paragraph{Code available.} The code we developed for our numerical simulations is available at the GitHub repository 
\\
\href{https://github.com/lanl/qubit-efficient-entanglement-spectroscopy}{\texttt{lanl/qubit-efficient-entanglement-spectroscopy}}

\paragraph{}
We use IBM's Qiskit~\cite{Qiskit} to perform our numerical tests. Qiskit is an open-source Python SDK for working with quantum circuits.
We implement our circuits using Qiskit and simulate them in the presence of noise using the QASM simulator from the Qiskit Aer module. All simulations were performed locally using Python version 3.6.9 and Qiskit version 0.12.1.

All circuits are implemented using a native gate set of $I, U_1, U_2$, and \CNOT{} and with operations Measurement and Reset, where
\[
    U_1(\lambda) := \begin{pmatrix} 1 & 0 \\ 0 & e^{i\lambda}\end{pmatrix}, \quad
    U_2(\phi,\lambda) := \frac{1}{\sqrt{2}}\begin{pmatrix} 1 & -e^{i \lambda} \\ e^{i\phi} & e^{i(\phi+\lambda)} \end{pmatrix}
\] 
are gates provided by Qiskit. Note that $H = U_2(0,\pi)$ and  $T = U_1(\pi/4)$. 
We decompose the \CSWAP{} gate as
\begin{equation*}
    \Qcircuit @C=0.5em @R=0.0em @!R {
        & \qw		& \qw 		& \qw		& \qw				& \ctrl{2}	& \qw		& \qw		& \qw				& \ctrl{2}	& \ctrl{1}	& \gate{T}			& \ctrl{1}	& \qw		& \qw \\
        & \targ		& \qw		& \ctrl{1}	& \qw				& \qwx\qw	& \qw		& \ctrl{1}	& \gate{T}			& \qwx\qw	& \targ		& \gate{T^\dagger}	& \targ		& \targ		& \qw \\
        & \ctrl{-1}	& \gate{H} 	& \targ		& \gate{T^{\dagger}}& \targ		& \gate{T}	& \targ		& \gate{T^\dagger}	& \targ		& \gate{T} 	& \gate{H}			& \qw		& \ctrl{-1}	& \qw \\
    }
\end{equation*}
where the top qubit is the control.

By default, Qiskit applies gates ``as soon as possible'', minimizing circuit depth by shifting gates to the left.
In order to correctly apply thermal noise, we insert identity gates to fill any gaps when a register must wait for operations to finish on other registers, taking into account the duration of each operation.
Thermal noise is applied on a gate-by-gate basis, but no gate noise, i.e. Pauli and depolarizing errors, is applied to the identity gates.
Other than the changes mentioned in this and the previous paragraph, all circuits are implemented as they appear in the figures.

The duration of each single-qubit gate is set to one timestep, the duration of a \CNOT{} gate five timesteps, the duration of a measurement three timesteps, and the duration of a qubit reset two timesteps (we always performed a measurement before performing a reset).

In all plots, each value of $\tr(\rho_A^n)$ is estimated using 100,000 runs.
For the plots which include the original \HT{} and \TCT{},
the probability of readout error is 2\%, which means that for each single-qubit measurement, there is a 2\% probability that the measurement result would be recorded incorrectly.
Thermal relaxation and decoherence errors are applied using parameters $T_1=T_2=2000$ and $T_{\text{pop}}=10^{-7}$. For an operation which takes time $t$,
let $p_\text{rel}=1-\exp(-t/T_1)$. Then, this means that for each qubit acted on, the probability that the qubit relaxes to $\ket{1}$ is $p_\text{rel} T_{\text{pop}}$ and the probability that the qubit relaxes to $\ket{0}$ is $p_\text{rel}\left(1-T_{\text{pop}}\right)$ (Qiskit can also apply a $Z$ operator to simulate decoherence, but for $T_1=T_2$, the probability of this is zero). 
A Pauli error channel is applied to all gates except identity such that for one-qubit gates, 
the probabilities of an $X,Y$, or $Z$ operator being applied are each $0.001$. 
A depolarizing error channel $E(\rho)=(1-\lambda)\rho + \lambda \tr(\rho)\frac{I}{2^m}$ is applied to all $m$-qubit gates except identity such that for single-qubit gates, $\lambda=0.001$.
For the \CNOT{} gate, the Pauli and depolarizing error parameters are multiplied by five.

For the plots which only include qubit-efficient algorithms, all of the noise parameters are set the same as above except for the Pauli and depolarization error parameters, which are reduced by a factor of ten.
The gate noise is reduced from the previous simulations in order to produce meaningful results for $n$ as high as twenty; we chose to reduce the gate noise because reducing the readout or thermal errors by a similar factor was not as effective.

All plots include error bars, although some of the bars may be too small to see.
The error bars in the plots of $\tr(\rho_A^n)$ versus experimental $\tr(\rho_A^n)$ are based on the expected statistical noise due to finite sampling and its effect on the post-processing formulas.
For the algorithms based on \HT{}, we use Hoeffding's inequality and a 68\% confidence level to calculate an additive error of at most $\pm 2\sqrt{-\ln(0.16)/(2S)}$, where $S$ is the number of trials performed.
For the algorithms based on \TCT{}, we calculate a confidence interval $[ c_{\text{low}},c_{\text{high}} ]$ for the raw output, $\abs{\bra{\Psi}M\ket{\Psi}}^2$, (before taking the square root) in the same way and set the final confidence interval to $[ \sqrt{c_{\text{low}}}, \sqrt{c_{\text{high}}}]$.
Note that unlike for \HT{}, the confidence intervals for \TCT{} are affected by $\tr(\rho_A^n)$, enlarging for smaller values. However, when  $\tr(\rho_A^n)$ is treated as a constant,
the size of the error bars scales as $O(1/\sqrt{S})$
in both cases.

The error bars in the plots of $n$ versus computed slopes are influenced by both statistical and simulated hardware noise.
The error bar for each point (a value of $n$ versus a slope) is calculated by applying a $t$-test with a 68\% confidence level to the linear regression which produced that slope. Intuitively, more linear underlying data produces smaller error bars.

\bibliographystyle{plainnat}

\bibliography{YirkaSubasi2021_arxiv_v2}

\end{document}